\documentclass[twocolumn]{aastex63}

\usepackage{CJK}                       
\usepackage[T1]{fontenc}               
\usepackage{amsmath,mathtools,tensor}  
\usepackage{color}                     
\usepackage{hyperref}                  

\hypersetup{colorlinks=true,linkcolor=black,citecolor=black,filecolor=black,urlcolor=black}

\graphicspath{{figures/}}

\allowdisplaybreaks[2]

\DeclarePairedDelimiter{\paren}{\lparen}{\rparen}
\DeclarePairedDelimiter{\bracket}{\lbrack}{\rbrack}
\DeclarePairedDelimiter{\braces}{\lbrace}{\rbrace}
\DeclarePairedDelimiter{\ave}{\langle}{\rangle}
\DeclarePairedDelimiter{\abs}{\lvert}{\rvert}
\newcommand{\parenpow}[3]{\paren[#1]{#2}^{\!#3}}

\DeclareMathOperator{\expm}{expm1}
\DeclareMathOperator{\diag}{diag}

\makeatletter
\newcommand{\tens}{\@ifstar\tensstar\tensnostar}
\makeatother
\newcommand{\tensstar}[2]{\tensor*{#1{\mathstrut}}{#2}}
\newcommand{\tensnostar}[2]{\tensor{#1{\mathstrut}}{#2}}

\newcommand{\dd}{\mathrm{d}}
\newcommand{\del}[1]{\tens{\nabla\!}{_{#1}}}
\newcommand{\pp}[1]{\tens{\partial}{_{#1}}}
\newcommand{\pfrac}[2]{\frac{\partial #1}{\partial #2}}

\newcommand{\ii}{\mathrm{i}}
\newcommand{\zeroh}{{\hat{0}}}
\newcommand{\zerob}{{\bar{0}}}
\newcommand{\oneh}{{\hat{1}}}
\newcommand{\oneb}{{\bar{1}}}
\newcommand{\twoh}{{\hat{2}}}
\newcommand{\twob}{{\bar{2}}}
\newcommand{\threeh}{{\hat{3}}}
\newcommand{\threeb}{{\bar{3}}}
\newcommand{\Eb}{\bar{E}}
\newcommand{\Ih}{{\hat{I}}}
\newcommand{\Ib}{{\bar{I}}}
\newcommand{\ah}{{\hat{a}}}
\newcommand{\ab}{{\bar{a}}}
\newcommand{\bh}{{\hat{b}}}
\newcommand{\bb}{{\bar{b}}}
\newcommand{\jh}{{\hat{\jmath}}}
\newcommand{\jb}{{\bar{\jmath}}}
\newcommand{\lh}{{\hat{l}}}
\newcommand{\mh}{{\hat{m}}}
\newcommand{\nh}{{\hat{n}}}
\newcommand{\sh}{{\hat{s}}}
\newcommand{\tb}{{\bar{t}}}
\newcommand{\xb}{{\bar{x}}}
\newcommand{\yb}{{\bar{y}}}
\newcommand{\zb}{{\bar{z}}}
\newcommand{\Omegah}{{\hat{\Omega}}}
\newcommand{\Omegab}{{\bar{\Omega}}}
\newcommand{\alphah}{{\hat{\alpha}}}
\newcommand{\alphab}{{\bar{\alpha}}}
\newcommand{\betah}{{\hat{\beta}}}
\newcommand{\betab}{{\bar{\beta}}}
\newcommand{\gammah}{{\hat{\gamma}}}
\newcommand{\gammab}{{\hat{\gamma}}}
\newcommand{\deltah}{{\hat{\delta}}}
\newcommand{\deltab}{{\bar{\delta}}}
\newcommand{\kappab}{{\bar{\kappa}}}
\newcommand{\zetah}{{\hat{\zeta}}}
\newcommand{\nuh}{{\hat{\nu}}}
\newcommand{\nub}{{\bar{\nu}}}
\newcommand{\psih}{{\hat{\psi}}}

\newcommand{\zerocon}[1]{\tens{0}{^{#1}}}
\newcommand{\zerocov}[1]{\tens{0}{_{#1}}}
\newcommand{\zerosplit}[2]{\tens*{0}{^{#1}_{#2}}}
\newcommand{\acon}[1]{\tens{A}{^{#1}}}

\newcommand{\xcon}[1]{\tens{x}{^{#1}}}
\newcommand{\etacon}[1]{\tens{\eta}{^{#1}}}
\newcommand{\etacov}[1]{\tens{\eta}{_{#1}}}
\newcommand{\gcon}[1]{\tens{g}{^{#1}}}
\newcommand{\gcov}[1]{\tens{g}{_{#1}}}
\newcommand{\betacon}[1]{\tens{\beta}{^{#1}}}
\newcommand{\betacov}[1]{\tens{\beta}{_{#1}}}
\newcommand{\gammacon}[1]{\tens{\gamma}{^{#1}}}
\newcommand{\gammacov}[1]{\tens{\gamma}{_{#1}}}
\newcommand{\ncon}[1]{\tens{\nh}{^{#1}}}
\newcommand{\ncov}[1]{\tens{\nh}{_{#1}}}
\newcommand{\naltconn}[1]{\tens*{n}{^{#1}_n}}
\newcommand{\kroncon}[1]{\tens{\delta}{^{#1}}}
\newcommand{\kronsplit}[2]{\tens*{\delta}{^{#1}_{#2}}}
\newcommand{\lorentz}[2]{\tens{\Lambda}{^{#1}_{#2}}}
\newcommand{\tet}[2]{\tens*{e}{^{#1}_{#2}}}
\newcommand{\rot}[2]{\tens*{\omega}{^{#1}_{#2}}}

\newcommand{\kcov}[1]{\tens{k}{_{#1}}}
\newcommand{\ucon}[1]{\tens{u}{^{#1}}}
\newcommand{\ucov}[1]{\tens{u}{_{#1}}}
\newcommand{\Bcon}[1]{\tens{B}{^{#1}}}
\newcommand{\Bcov}[1]{\tens{B}{_{#1}}}
\newcommand{\bcon}[1]{\tens{b}{^{#1}}}
\newcommand{\bcov}[1]{\tens{b}{_{#1}}}
\newcommand{\ugascon}[1]{\tens*{u}{_{\mathrm{gas}}^{#1}}}
\newcommand{\uradcon}[1]{\tens*{u}{_{\mathrm{rad}}^{#1}}}
\newcommand{\vradcon}[1]{\tens*{v}{_{\mathrm{rad}}^{#1}}}
\newcommand{\Rcon}[1]{\tens{R}{^{#1}}}
\newcommand{\Rcone}[1]{\tens*{R}{_{\mathrm{exact}}^{#1}}}
\newcommand{\Rconex}[1]{\tens*{R}{_{\mathrm{exact},x}^{#1}}}
\newcommand{\Rsplit}[2]{\tens{R}{^{#1}_{#2}}}
\newcommand{\Tcon}[1]{\tens{T}{^{#1}}}
\newcommand{\Tsplit}[2]{\tens{T}{^{#1}_{#2}}}
\newcommand{\Tgascon}[1]{\tens*{T}{_{\mathrm{gas}}^{#1}}}
\newcommand{\Tradcon}[1]{\tens*{T}{_{\mathrm{rad}}^{#1}}}
\newcommand{\Fstarcon}[1]{\tens{(\star F)}{^{#1}}}
\newcommand{\Gcon}[1]{\tens{G}{^{#1}}}
\newcommand{\Gcov}[1]{\tens{G}{_{#1}}}
\newcommand{\Fcon}[1]{\tens{F}{^{#1}}}
\newcommand{\Fcov}[1]{\tens{F}{_{#1}}}

\newcommand{\cm}{\mathrm{cm}}
\newcommand{\g}{\mathrm{g}}
\newcommand{\erg}{\mathrm{erg}}
\newcommand{\K}{\mathrm{K}}

\newcommand{\arad}{a_\mathrm{rad}}
\newcommand{\aradzero}{a_{\mathrm{rad},0}}
\newcommand{\aradbar}{\bar{a}_\mathrm{rad}}
\newcommand{\mprot}{m_\mathrm{p}}
\newcommand{\kB}{k_\mathrm{B}}
\newcommand{\Msun}{M_\odot}

\newcommand{\Ihl}{\Ih_\mathrm{L}}
\newcommand{\Ihr}{\Ih_\mathrm{R}}
\newcommand{\Epeak}{E_\mathrm{peak}}
\newcommand{\Eoff}{E_\mathrm{off}}
\newcommand{\jba}{\jb^\mathrm{a}}
\newcommand{\jbs}{\jb^\mathrm{s}}
\newcommand{\alphaea}{\alpha_\mathrm{E}^\mathrm{a}}
\newcommand{\alphaes}{\alpha_\mathrm{E}^\mathrm{s}}
\newcommand{\alphahe}{\alphah_\mathrm{E}}
\newcommand{\alphaba}{\alphab^\mathrm{a}}
\newcommand{\alphabs}{\alphab^\mathrm{s}}
\newcommand{\alphabe}{\alphab_\mathrm{E}}
\newcommand{\alphabea}{\alphab_\mathrm{E}^\mathrm{a}}
\newcommand{\alphabes}{\alphab_\mathrm{E}^\mathrm{s}}

\newcommand{\alphabpa}{\alphab_\mathrm{P}^\mathrm{a}}
\newcommand{\alphabf}{\alphab_\mathrm{F}}
\newcommand{\kappaa}{\kappa^\mathrm{a}}
\newcommand{\kappas}{\kappa^\mathrm{s}}
\newcommand{\kappaba}{\kappab^\mathrm{a}}
\newcommand{\nzerocrit}{\nh_0^\mathrm{crit}}
\newcommand{\Gammagas}{\Gamma_\mathrm{gas}}
\newcommand{\prad}{p_\mathrm{rad}}
\newcommand{\pradzero}{p_{\mathrm{rad},0}}
\newcommand{\pradbar}{\bar{p}_\mathrm{rad}}
\newcommand{\pgas}{p_\mathrm{gas}}
\newcommand{\pgaszero}{p_{\mathrm{gas},0}}
\newcommand{\pgasbar}{\bar{p}_\mathrm{gas}}
\newcommand{\pgaspeakbar}{\bar{p}_\mathrm{gas,peak}}
\newcommand{\pmag}{p_\mathrm{mag}}
\newcommand{\ptot}{p_\mathrm{tot}}
\newcommand{\ptotzero}{p_{\mathrm{tot},0}}
\newcommand{\urad}{u_\mathrm{rad}}
\newcommand{\uradzero}{u_{\mathrm{rad},0}}
\newcommand{\uradbar}{\bar{u}_\mathrm{rad}}
\newcommand{\ugas}{u_\mathrm{gas}}
\newcommand{\ugaszero}{u_{\mathrm{gas},0}}
\newcommand{\ugasbar}{\bar{u}_\mathrm{gas}}
\newcommand{\umag}{u_\mathrm{mag}}
\newcommand{\utot}{u_\mathrm{tot}}
\newcommand{\wrad}{w_\mathrm{rad}}
\newcommand{\wgas}{w_\mathrm{gas}}
\newcommand{\wmag}{w_\mathrm{mag}}
\newcommand{\Tradzero}{T_{\mathrm{rad},0}}
\newcommand{\Tgaszero}{T_{\mathrm{gas},0}}
\newcommand{\Tequil}{T_\mathrm{equil}}
\newcommand{\Tinf}{T_\infty}
\newcommand{\Tpeakbar}{\bar{T}_\mathrm{peak}}
\newcommand{\cs}{c_\mathrm{s}}
\newcommand{\ca}{c_\mathrm{a}}
\newcommand{\rg}{r_\mathrm{g}}
\newcommand{\tg}{t_\mathrm{g}}
\newcommand{\redgebar}{\bar{r}_\mathrm{edge}}
\newcommand{\rpeakbar}{\bar{r}_\mathrm{peak}}
\newcommand{\rhopeak}{\rho_\mathrm{peak}}
\newcommand{\rhopeakbar}{\bar{\rho}_\mathrm{peak}}

\newcommand{\xmin}{x_\mathrm{min}}
\newcommand{\xmax}{x_\mathrm{max}}
\newcommand{\Nang}{N_\mathrm{ang}}
\newcommand{\Nlev}{N_\mathrm{lev}}
\newcommand{\Nsamp}{N_\mathrm{samp}}
\newcommand{\Ncell}{N_\mathrm{cell}}
\newcommand{\Nprim}{N_\mathrm{prim}}
\newcommand{\prim}{\mathcal{P}}
\newcommand{\cons}{\mathcal{C}}
\newcommand{\flux}{\mathcal{F}}
\newcommand{\source}{\mathcal{S}}

\newcommand{\code}[1]{\texttt{#1}}

\definecolor{electricblue}{RGB}{6,82,255}

\shorttitle{Athena++ Radiation--GRMHD}
\shortauthors{White et al.}

\begin{document}
\begin{CJK*}{UTF8}{gbsn}

\title{An Extension of the Athena++ Code Framework for Radiation--Magnetohydrodynamics \\ in General Relativity Using a Finite-Solid-Angle Discretization}
\author{Christopher J. White}
\affiliation{Center for Computational Astrophysics, Flatiron Institute, New York, NY}
\affiliation{Department of Astrophysical Sciences, Princeton University, Princeton, NJ}
\author{Patrick D. Mullen}
\affiliation{CCS-2, Los Alamos National Laboratory, Los Alamos, NM}
\affiliation{Center for Theoretical Astrophysics, Los Alamos National Laboratory, Los Alamos, NM}
\author{Yan-Fei Jiang (姜燕飞)}
\affiliation{Center for Computational Astrophysics, Flatiron Institute, New York, NY}
\author{Shane W. Davis}
\affiliation{Department of Astronomy, University of Virginia, Charlottesville, VA}
\author{James M. Stone}
\affiliation{School of Natural Sciences, Institute for Advanced Study, Princeton, NJ}
\author{Viktoriya Morozova}
\affiliation{Institute for Gravitation and the Cosmos, The Pennsylvania State University, University Park, PA}
\author{Lizhong Zhang (张力中)}
\affiliation{Department of Physics, University of California, Santa Barbara, CA}

\begin{abstract}

We extend the general-relativistic magnetohydrodynamics (GRMHD) capabilities of \code{Athena++} to incorporate radiation. The intensity field in each finite-volume cell is discretized in angle, with explicit transport in both space and angle properly accounting for the effects of gravity on null geodesics, and with matter and radiation coupled in a locally implicit fashion. Here we describe the numerical procedure in detail, verifying its correctness with a suite of tests. Motivated in particular by black hole accretion in the high-accretion-rate, thin-disk regime, we demonstrate the application of the method to this problem. With excellent scaling on flagship computing clusters, the port of the algorithm to the GPU-enabled \code{AthenaK} code now allows the simulation of many previously intractable radiation--GRMHD systems.

\strut

\end{abstract}

\section{Introduction}
\label{sec:introduction}

There are a number of astrophysical systems in which the effects of both general relativity (GR) and radiation in the form of photons or neutrinos significantly modify the hydrodynamics (HD) or magnetohydrodynamics (MHD) of the fluid that is present. Black holes accreting at sufficiently high rates, merging neutron stars, and the innermost regions of core-collapse supernovae all belong to this category. These systems are turbulent and susceptible to intrinsically multidimensional fluid instabilities and phenomena \citep{Abramowicz1988,Done2007,McKinney2014,Faber2012,Baiotti2017,Metzger2020,Nordhaus2010,Janka2012,Muller2016}, and so modeling and understanding them calls for accurate simulations that solve the governing differential equations.

Radiation in particular is difficult to model well. In the continuum limit, collisional MHD fluids are described by a small number (${\sim}8$) of dynamical values at every point in space; spatial discretization results in a finite count of real numbers to evolve in time. Radiation, however, is not collisional, and so even limiting the calculation to a single frequency, the dynamical quantity comprises a functional degree of freedom at every point in space.\footnote{In particular, continuum radiation is described by nonnegative functions on the sphere, which need not be smooth.} That is, the radiation field consists of intensity that depends on three-dimensional spatial location and two-dimensional direction. With frequencies considered, this amounts to solving the Boltzmann equation on a six-dimensional domain.

Radiation methods can be categorized based on how they reduce this complexity to a more tractable level.\footnote{More mathematically, methods can be distinguished by how the radiation field at a point is approximated with only a finite count of real numbers.} It is conceptually straightforward to divide the sphere into discrete solid angles just as finite-volume fluid methods discretize space, evolving some average representation of the intensity within each solid angle and spatial cell. This approach, called a Boltzmann, discrete-ordinate \citep{Krook1955}, or $S_N$ \citep{Grant1968} method in various contexts, is the one we adopt here. It has a history dating back to \citet{Chandrasekhar1950}, and it is conducive to converging to the correct solution \citep{Madsen1971}.

Values on a sphere can instead be represented by spherical harmonics. These are evolved in the $P_N$ method, which has been applied to general-relativistic transport by \citet{Radice2013}. Conceptually intermediate between $S_N$ and $P_N$, finite-element discretizations of the sphere have been considered for special-relativistic transport \citep{Bhattacharyya2022}, with the GR case still under development.

An alternative strategy, and one with long historical precedent in astrophysics, is to describe the radiation field at a point with a finite number of moments. As the evolution of the $k$-th moment depends only on moment $k + 1$, a closure relation that prescribes the $n$-th moment as a function of moments $0$ through $n - 1$ truncates the series. This applies to relativistic fluids \citep{Thorne1981} as well as it does to Newtonian fluids. Flux-limited diffusion sets $n = 1$ and allows spatial gradients in the zeroth moment, modified by a flux limiter, to contribute to the closure. Commonly, $n$ is chosen to be $2$, and the different assumed prescriptions lead to different flavors of the M1 method \citep{Levermore1984}. The accuracy of this strategy hinges on the validity of the assumed closure. Unfortunately, it is quite possible for different parts of a physical system to have similar zeroth and first moments but very different second moments, and for some problems no amount of computational power can overcome the fact that the span of radiation fields at a point representable with M1 may be insufficient to capture this diversity. Some shortcomings of M1 closures in a realistic GRMHD accretion setting are shown in \citet{Asahina2022}.

The aforementioned discrete-ordinate method is more conducive to converging to the correct solution with increased computational power, as the sphere can simply be divided into finer solid angles. Another alternative is to use Monte Carlo sampling of the radiation field, launching appropriately distributed photon packets to propagate though space according to the geodesic equation \citep{Fleck1971}. This method converges to the correct solution with sufficiently many samples, though optically thick regions require careful treatment to avoid prohibitive costs. Moreover, parallelization and load balancing of large problems is difficult in this case: Domain decomposition only in space can lead to drastically different computational costs for different chunks of the domain based on how many packets are needed, while distributing only the packets may require prohibitively large amounts of memory for each process to store the fluid state of the entire system.

Despite the difficulties, recent years have seen much progress in unifying GR, radiation, and (M)HD in astrophysical codes. Within the core-collapse community, it is common for relativistic dynamics, strong gravity, and neutrino transport to be combined in an M1 method that is simplified by leveraging the approximately spherical nature of the system or the expected approximate axisymmetry of the radiation field at every point about the radial direction \citep{Liebendorfer2004,OConnor2015}. More multidimensional relativistic M1 schemes have been presented in \citet{Farris2008}, \citet{Shibata2011}, \citet{Zanotti2011}, \citet{Roedig2012}, \citet{Sadowski2013}, \citet{McKinney2014}, \citet{Fragile2014}, \citet{Foucart2015}, \citet{Kuroda2016}, \citet{Takahashi2016}, \citet{Rahman2019}, \citet{Weih2020}, \citet{Anninos2020}, \citet{Radice2022}, and \citet{Cheong2023}.

Monte Carlo sampling has been employed in GRMHD by \citet{Ryan2015}, \citet{Foucart2021}, \citet{Roth2022}, and \citet{Kawaguchi2023}. An example of a hybrid scheme is that of \citet{Foucart2018}, where Monte Carlo techniques are used to sample the local radiation field in order to provide more situationally appropriate closures for a moment method. More deterministically, a method-of-characteristics approach is used to close the moment equations in the MOCMC scheme of \citet{Ryan2020}. Similarly, Boltzmann solvers have been used to inform moment closures, both in spherically symmetric codes \citep{Muller2010,Roberts2012} and in multidimensional contexts \citep{Asahina2020}.

More scarce in the study of accretion disks are pure, deterministic Boltzmann solvers that treat the discretized intensity as a function of angle as fundamental. The codes developed by \citet{Nagakura2017} and \citet{Chan2020} do this, but they are both designed for the core-collapse problem, focusing on the complexities of multi-species, multi-energy-group transport and coupling while neglecting magnetic fields.

Here we describe our implementation of general-relativistic radiation, solving the GR Boltzmann transport equation as given, for example, in \citet{Shibata2014} and especially \citet{Davis2020}, coupled to a relativistic MHD fluid within the \code{Athena++} framework \citep{Stone2020}. The MHD capabilities of the code in special relativity (SR) and stationary-spacetime GR \citep{White2016} have been used for a number of applications. At the same time, the Newtonian MHD sector of the framework has been applied to a number of radiative problems with a variety of methods; in particular, \citet{Jiang2014} develops the explicit Boltzmann transport whose relativistic generalization is our current focus. Coupling of flat-spacetime SRMHD with radiation in \code{Athena++} has been developed and employed by \citet{Zhang2021}, and so we now turn to the GR case.

The method described here allows a number of problems at the forefront of astrophysical modeling---in particular, black holes accreting near or beyond the Eddington rate---to be simulated in a manner that provides for convergence to the physical solution even with complex radiation fields. While our method supports multiple frequency groups in a straightforward way, we will consider mostly gray transport here, leaving frequency-dependent tests and applications for a future work. At the same time, this generalization of \citet{Jiang2014} to account for gravitational bending of light naturally allows the coupling of radiation to Newtonian or SR fluids in more 1D and 2D geometries than previously supported.

Our numerical implementation is detailed in Section~\ref{sec:method}. Section~\ref{sec:tests} demonstrates the code passing a variety of tests of both transport and radiation--fluid coupling in both flat and curved spacetimes. We showcase a 3D simulation of black hole accretion in Section~\ref{sec:example}. Section~\ref{sec:performance} provides performance statistics for the code, especially using the new performance-portable \code{AthenaK} version of the framework (Stone et al., in prep.). A summary of capability and future outlook is provided in Section~\ref{sec:summary}.

We will generally omit explicit factors of $c$ and $G$, and we will employ metrics with the $(-,\, +,\, +,\, +)$ signature. In what follows, our symmetrization convention omits combinatorial factors. For example, we have
\begin{subequations} \begin{align}
  \acon{(\alpha\beta)} & = \acon{\alpha\beta} + \acon{\beta\alpha}, \\
  \acon{[\alpha\beta]} & = \acon{\alpha\beta} - \acon{\beta\alpha}.
\end{align} \end{subequations}

\section{Numerical Method}
\label{sec:method}

The crux of our algorithm is the time evolution of the specific intensity of radiation, $\Ih_\nu$, which is a function of three spatial dimensions and three momentum directions (one frequency or equivalently energy $\nuh$, a polar angle $\zetah$, and an azimuthal angle $\psih$). From \citet{Davis2020} we can write
\begin{widetext}
\begin{equation} \label{eq:radiation_differential}
  \del{\alpha} \paren[\big]{\ncon{\alpha} \ncov{\beta} \Ih_\nu} + \pp{\nuh} \paren[\big]{\ncon{\nuh} \ncov{\beta} \Ih_\nu} + \sh^{-1} \pp{\zetah} \paren[\big]{\sh \ncon{\zetah} \ncov{\beta} \Ih_\nu} + \pp{\psih} \paren[\big]{\ncon{\psih} \ncov{\beta} \Ih_\nu} = \ncov{\beta} (\jh_\nu - \alphah_\nu \Ih_\nu),
\end{equation}
\end{widetext}
The first term, being a covariant divergence, encodes both the time derivative of our conserved quantity $\nh^0 \nh_\beta \Ih_\nu$ as well as spatial derivatives of spatial fluxes $\nh^a \nh_\beta \Ih_\nu$. The second term yields gravitational redshifting, while the third and fourth terms account for light changing direction as it moves through space. Here, $\jh_\nu$ and $\alphah_\nu$ are the emission and absorption coefficients coupling radiation to any fluid present, the components of $\nh$ are position-~and momentum-dependent geometrical factors explained below, and we use the shorthand
\begin{equation}
  \sh = \sin\zetah.
\end{equation}

\begin{figure*}
  \centering
  \includegraphics{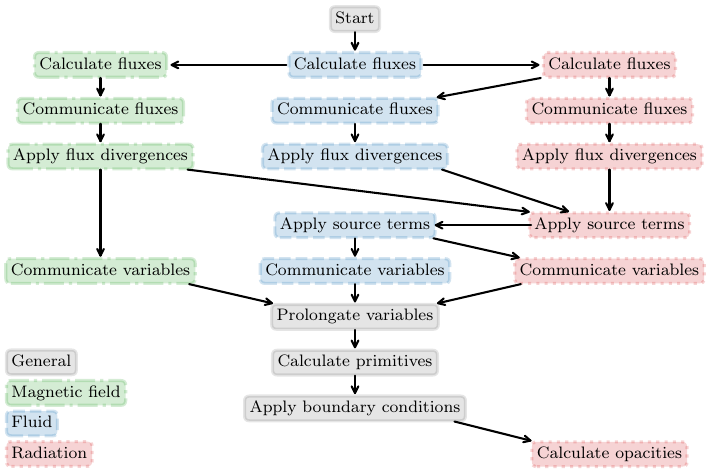}
  \caption{Tasks and their dependencies for a single time-integration stage. A task cannot begin until all tasks pointing to it have successfully terminated. Our finite-volume formulation of radiation is conceptually similar to that of hydrodynamics and magnetic field evolution, and so the sets of tasks associated to these overarching parts of the code largely parallel one another. \label{fig:tasks}}
\end{figure*}

After defining the coordinates we use (\S\ref{sec:method:coordinates}) and deriving the discretized forms of the equations we evolve (\S\ref{sec:method:equations}), we will detail the principal components of our time integration algorithm for radiation. As with finite-volume hydrodynamics, the form of the differential equation defines the aforementioned conserved quantity and spatial fluxes, as well as fluxes in frequency and angle, as the terms to which partial derivatives are applied. We can take specific intensity to be the primitive variable. Paralleling the evolution of the fluid and magnetic field, each stage in a multi-stage time integrator performs the following tasks:
\begin{enumerate}
  \item Calculate fluxes: Intensities are reconstructed (interpolated) from cell centers to faces in both space and angle (\S\ref{sec:method:reconstruction}). Fluxes are then calculated by upwinding reconstructed values, while being aware of optical depths (\S\ref{sec:method:fluxes}).
  \item Communicate fluxes: Fluxes are synchronized across coarse/fine mesh-refinement boundaries, with coarse values overwritten by any finer values obtained at the same spatial coordinates.
  \item Apply flux divergences: Conserved values are updated via the appropriate divergences of spatial and angular fluxes.
  \item Apply source terms: Conserved values are locally implicitly updated in accordance with the inhomogeneous terms in the equations stemming from matter--radiation coupling (\S\ref{sec:method:coupling}).
  \item Communicate variables: Conserved values from active zones are sent to their corresponding ghost zones. In the case of fine-to-coarse communication across a refinement boundary, the values are restricted before communication.
  \item Prolongate variables: Ghost-zone values received in coarse-to-fine communication are prolongated by the recipients.
  \item Calculate primitives: The primitive intensities are calculated from their conserved counterparts.
  \item Apply boundary conditions: Ghost zones external to the physical domain are populated based on the desired behavior at the boundaries.
\end{enumerate}
Figure~\ref{fig:tasks} shows the tasks involved in a radiation-MHD calculation with mesh refinement. Arrows indicate dependencies; a task cannot commence until all tasks pointing to it have completed.

Further details regarding angular discretization are given in Section~\ref{sec:method:angular_grids}.

\subsection{Coordinates}
\label{sec:method:coordinates}

As we will make use of no fewer than four coordinate systems, we are careful to distinguish them with consistent notation throughout this work.

Spacetime is parameterized by coordinates $x^\alpha$, indexed with $\alpha$, $\beta$, etc. Unspecified spatial components will be indexed with $a$, $b$, etc. We will refer to tensors with components in this frame as being in the ``coordinate frame.'' We will only consider stationary (but certainly not necessarily static) spacetimes here, subject to
\begin{equation} \label{eq:stationary_metric}
  \pp{0} \gcov{\alpha\beta} = 0.
\end{equation}

At times, we will refer to quantities in the ``normal frame,'' indicated by primes on the indices (e.g., $x^{\alpha'}$).\footnote{This frame is often indicated by tildes on the tensor itself (e.g., $\tilde{u}^i$) in the GRMHD literature \citep[e.g.,][]{Gammie2003}.} This is the frame in which the time direction is made orthogonal to surfaces of constant coordinate time. In terms of the standard $3{+}1$ lapse, shift, and spatial metric
\begin{subequations} \begin{align}
  \alpha & = \parenpow{\big}{-\gcon{00}}{-1/2}, \\
  \betacov{a} & = \gcov{0a}, \\
  \betacon{a} & = \alpha^2 \gcon{0a}, \\
  \gammacov{ab} & = \gcov{ab}, \\
  \gammacon{ab} & = \alpha^2 \paren[\big]{\gcon{0a} \gcon{0b} - \gcon{00} \gcon{ab}},
\end{align} \end{subequations}
the metric in this frame is given by
\begin{subequations} \begin{align}
  \gcov{0'0'} & = -1, & \gcon{0'0'} & = -1, \\
  \gcov{0'a'} & = 0, & \gcon{0'a'} & = 0, \\
  \gcov{a'b'} & = \gammacov{ab}, & \gcon{a'b'} & = \gammacon{ab}.
\end{align} \end{subequations}
Vectors can be transformed between the coordinate and normal frames via
\begin{subequations} \begin{align}
  \acon{0'} & = \alpha \acon{0}, \\
  \acon{a'} & = \acon{a} + \betacon{a} \acon{0},
\end{align} \end{subequations}
with the rules for higher-rank tensors naturally following from these.

We will also rely heavily on an orthonormal frame, which we will call the ``tetrad frame.'' Indices in this frame will be decorated with circumflexes (e.g., $x^\alphah$). The tetrad frame is related to the coordinate frame at any location via the set of $16$ spatially dependent components $e^\alpha_\betah$:
\begin{subequations} \label{eq:tetrad} \begin{align}
  \xcon{\alpha} & = \tet{\alpha}{\betah} \xcon{\betah}, \label{eq:tetrad:forward} \\
  \xcon{\alphah} & = \tet{\alphah}{\beta} \xcon{\beta}.
\end{align} \end{subequations}
Indices can be raised or lowered with the appropriate metric, whether Minkowski or the spacetime metric at hand:
\begin{equation}
  \tet{\alphah}{\beta} = \etacon{\alphah\gammah} \gcov{\beta\delta} \tet{\delta}{\gammah}.
\end{equation}
The components form a valid tetrad only if
\begin{equation}
  \gcov{\alpha\beta} \tet{\alpha}{\gammah} \tet{\beta}{\deltah} = \etacov{\gammah\deltah},
\end{equation}
which fixes $10$ degrees of freedom at each point in spacetime. We further require that the timelike direction in the tetrad frame be orthogonal to surfaces of constant coordinate time $x^0$, and that this be a future-pointing direction. Mathematically, this fixes another $3$ degrees of freedom and forces
\begin{equation}
  \tet{\alpha}{\zeroh} = -\alpha \gcon{0\alpha}.
\end{equation}
The remaining $3$ degrees of freedom can be leveraged to increase numerical accuracy in problem-dependent ways. See Section~\ref{sec:method:angular_grids} for more details regarding angular grids, including geodesic grids supported by the code. As with the coordinate frame, we will restrict attention to tetrads that are stationary:
\begin{equation} \label{eq:stationary_tetrad}
  \pp{0} \tet{\alpha}{\betah} = 0.
\end{equation}

\begin{figure}
  \centering
  \includegraphics{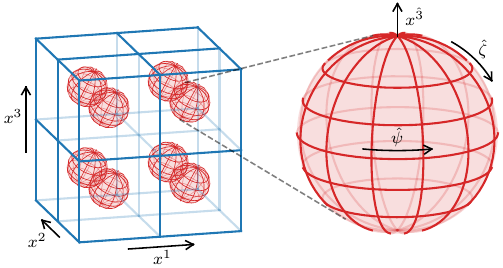}
  \caption{Schematic illustrating the relationship between spatial coordinates $x^a$ and angular directions $\zetah$ and $\psih$. Space is divided into cells by surfaces of constant $x^a$. Each cell has associated with it a grid of angles, separated by arcs of constant $\zetah$ and $\psih$. The relative orientations of the two grids are determined by the spatially dependent tetrad, which relates $x^\threeh$, as well as $x^\oneh$ and $x^\twoh$ (not shown), to $x^1$, $x^2$, and $x^3$. \label{fig:coordinates}}
\end{figure}

The tetrad frame can be used to define spherical coordinates characterizing momentum at every point in spacetime. Following \citet{Davis2020}, define $\zetah$ to be the polar angle measured from the $+x^{\threeh}$-direction in the tangent space at the point in question, and let $\psih$ be the azimuthal angle measured in the usual sense from the $+x^{\oneh}$-direction. Figure~\ref{fig:coordinates} illustrates how these coordinates parameterize direction within every cell on a spatial grid. The frequency $\nuh$ in this frame characterizes the magnitude of momentum. Angles $\zetah$ and $\psih$ will be indexed by $l$ and $m$. The naturally arising latitude/longitude grid is discussed further in Section~\ref{sec:method:angular_grids}, together with an alternative geodesic grid for radiation angles.

Finally, we will use macrons (e.g., $x^\alphab$) to denote components in a particular orthonormal rest frame of the fluid. Given the fluid's velocity in the tetrad frame, we define this frame via
\begin{equation}
  \acon{\alphab} = \lorentz{\alphab}{\betah} \acon{\betah},
\end{equation}
with
\begin{subequations} \begin{align}
  \lorentz{\zerob}{\zeroh} & = \ucon{\zeroh}, \\
  \lorentz{\zerob}{\ah} & = -\ucov{\ah}, \\
  \lorentz{\ab}{\zeroh} & = -\ucon{\ah}, \\
  \lorentz{\ab}{\bh} & = \frac{\ucon{\ah} \ucov{\bh}}{1 + u^\zeroh} + \kronsplit{a}{b}.
\end{align} \end{subequations}

\subsection{Governing Equations}
\label{sec:method:equations}

Let $\Ih_\nu$ be the tetrad-frame specific intensity, defined in general as a function of $x^\alpha$, $\nuh$, $\zetah$, and $\psih$. Let $\nh$ be the future-directed null vector aligned with momentum that has spatial length unity in the tetrad frame. That is, $\nh^\alphah$ is a function of $\zetah$ and $\psih$ with components
\begin{subequations} \begin{align}
  \ncon{\zeroh} & = 1, \\
  \ncon{\oneh} & = \sin\zetah \cos\psih, \\
  \ncon{\twoh} & = \sin\zetah \sin\psih, \\
  \ncon{\threeh} & = \cos\zetah.
\end{align} \end{subequations}
Note that, while $\nh^\alphah$ is only a function of direction coordinates $\zetah$ and $\psih$, $\nh^\alpha$ also depends on position $x^a$ via the spatially dependent tetrad (cf.\ Equation~\eqref{eq:tetrad:forward}). The latter, spatially varying components appear in most equations, such as the original Equation~\eqref{eq:radiation_differential}.

Additionally, define the symbols
\begin{subequations} \begin{align}
  \ncon{\nuh} & = -\nuh \ncon{\alphah} \ncon{\betah} \rot{\zeroh}{\alphah\betah}, \\
  \ncon{\zetah} & = \sh^{-1} \ncon{\alphah} \ncon{\betah} \ncon{[\zeroh} \rot{\threeh]}{\alphah\betah}, \\
  \ncon{\psih} & = \sh^{-2} \ncon{\alphah} \ncon{\betah} \ncon{[\twoh} \rot{\oneh]}{\alphah\betah},
\end{align} \end{subequations}
where
\begin{equation} \label{eq:ricci}
  \rot{\gammah}{\alphah\betah} = \tet{\gammah}{\epsilon} \tet{\delta}{\betah} \del{\delta} \tet{\epsilon}{\alphah}
\end{equation}
are the Ricci rotation coefficients. Given our stationary assumptions Equation~\eqref{eq:stationary_metric} and~\eqref{eq:stationary_tetrad}, these three symbols, as well as $\nh^\alpha$ and $\nh^\alphah$, need only be computed once during a simulation. With these definitions, the differential equation of general-relativistic radiative transport from \citet{Davis2020} can be written as Equation~\eqref{eq:radiation_differential}.

The covariant $\beta$ in Equation~\eqref{eq:radiation_differential} can be any spacetime index, and its choice will govern what quantity will be exactly conserved by the numerical scheme. For example, if $\{x^\alpha\}$ describes axisymmetric coordinates, then setting $\beta = \phi$ will conserve a form of angular momentum. For this work we choose to let $\beta$ be the time coordinate, conserving the covariant radiation energy density exactly.

We must also solve the equations of GRMHD, modified to account for four-momentum transfer between the fluid and radiation:
\begin{subequations} \label{eq:grmhd_differential} \begin{align}
  \del{\alpha} \paren[\big]{\rho \ucon{\alpha}} & = 0, \label{eq:grmhd_differential:mass} \\
  \del{\alpha} \Tsplit{\alpha}{\beta} & = \Gcov{\beta}, \label{eq:grmhd_differential:momentum} \\
  \del{\alpha} \Fstarcon{\beta\alpha} & = 0,
\end{align} \end{subequations}
where $\rho$ is the fluid rest-mass density, $u$ is its four-velocity, $T$ is the stress--energy tensor of the fluid and electromagnetic field, $F$ is the electromagnetic field tensor, and $\star$ is the Hodge dual operator. Formally, the coupling term is given by
\begin{equation} \label{eq:coupling_force}
  \Gcov{\alpha} = -\oint \! \int_0^\infty \ncov{\alpha} (\jh_\nu - \alphah_\nu \Ih_\nu) \, \dd\nuh \, \dd\Omegah,
\end{equation}
where $\dd\Omegah = \sh \, \dd\zetah \, \dd\psih$ is the standard spherical area element in the tetrad frame. However, as described in Section~\ref{sec:method:coupling}, this term is not directly computed from this formula.

Just as Equations~\eqref{eq:grmhd_differential} are averaged over a spacetime volume element to yield the finite-volume form of the equations solved by \code{Athena++}, so too is Equation~\eqref{eq:radiation_differential} averaged over spacetime volume, frequency interval, and direction solid angle. We employ the notation
\begin{widetext}
\begin{subequations} \begin{align}
  \Delta t & = \int \, \dd\xcon{0}, & \ave{\cdot}_t & = \frac{1}{\Delta t} \int \! \mathrel{\cdot} \, \dd t, \\
  \Delta V & = \int \sqrt{-g} \, \dd\xcon{1} \, \dd\xcon{2} \, \dd\xcon{3}, & \ave{\cdot}_V & = \frac{1}{\Delta V} \int \! \mathrel{\cdot} \sqrt{-g} \, \dd\xcon{1} \, \dd\xcon{2} \, \dd\xcon{3}, \\
  \Delta A_a & = \int \sqrt{-g} \, \dd\xcon{b} \, \dd\xcon{c}, & \ave{\cdot}_{A_a} & = \frac{1}{\Delta A_a} \int \! \mathrel{\cdot} \sqrt{-g} \, \dd\xcon{b} \, \dd\xcon{c}, \\
  \Delta\nuh & = \int \, \dd\nuh, & \ave{\cdot}_\nuh & = \frac{1}{\Delta\nuh} \int \! \mathrel{\cdot} \, \dd\nuh, \\
  \Delta\Omegah & = \int \sh \, \dd\zetah \, \dd\psih, & \ave{\cdot}_\Omegah & = \frac{1}{\Delta\Omegah} \int \! \mathrel{\cdot} \sh \, \dd\zetah \, \dd\psih, \\
  \Delta\lambda_\lh & = \int \sh \, \dd\lambda^\mh, & \ave{\cdot}_{\lambda_\lh} & = \frac{1}{\Delta\lambda_\lh} \int \! \mathrel{\cdot} \sh \, \dd\lambda^\mh,
\end{align} \end{subequations}
where $a,\, b,\, c \in \{1,\, 2,\, 3\}$ are always distinct, as are $l,\, m \in \{\zeta,\, \psi\}$, with $\lambda^\zetah = \zetah$ and $\lambda^\psih = \psih$. Then the integral form of the radiative transfer equation can be written\footnote{We omit the term $\ave{(\partial_0 g_{\alpha\beta}) \nh^\alpha \nh^\beta \Ih_\nu}_{t V \nuh \Omegah} / 2$ that is added to the right-hand side but vanishes given Equation~\eqref{eq:stationary_metric}.}
\begin{multline} \label{eq:radiation_integral_frequency}
  \frac{1}{\Delta t} \bracket[\Big]{\ave[\big]{\ncon{0} \ncov{0} \Ih_\nu}_{V \nuh \Omegah}}_{x^0_-}^{x^0_+} + \sum_a \frac{1}{\Delta V} \bracket[\Big]{\Delta A_a \ave[\big]{\ncon{a} \ncov{0} \Ih_\nu}_{t A_a \nuh \Omegah}}_{x^a_-}^{x^a_+} + \frac{1}{\Delta\nuh} \bracket[\Big]{\ave[\big]{\ncon{\nuh} \ncov{0} \Ih_\nu}_{t V \Omegah}}_{\nuh_-}^{\nuh_+} \\ + \sum_\lh \frac{1}{\Delta\Omegah} \bracket[\Big]{\Delta\lambda_\lh \ave[\big]{\ncon{\lh} \ncov{0} \Ih_\nu}_{t V \nuh \lambda_\lh}}_{\lambda^\lh_-}^{\lambda^\lh_+} = \ave[\big]{\ncov{0} (\jh_\nu - \alphah_\nu \Ih_\nu)}_{t V \nuh \Omegah}.
\end{multline}
Specializing to the case of gray transport, we can consider frequency to have a single bin $0 < \nuh < \infty$, in which case the equation can be written
\begin{equation} \label{eq:radiation_integral}
  \frac{1}{\Delta t} \bracket[\Big]{\ave[\big]{\ncon{0} \ncov{0} \Ih}_{V \Omegah}}_{x^0_-}^{x^0_+} + \sum_a \frac{1}{\Delta V} \bracket[\Big]{\Delta A_a \ave[\big]{\ncon{a} \ncov{0} \Ih}_{t A_a \Omegah}}_{x^a_-}^{x^a_+} \\ + \sum_\lh \frac{1}{\Delta\Omegah} \bracket[\Big]{\Delta\lambda_\lh \ave[\big]{\ncon{\lh} \ncov{0} \Ih}_{t V \lambda_\lh}}_{\lambda^\lh_-}^{\lambda^\lh_+} = \ave[\big]{\ncov{0} (\jh - \alphahe \Ih)}_{t V \Omegah}.
\end{equation}
\end{widetext}
Here we have defined the gray intensity, gray emissivity, and energy-mean absorptivity
\begin{subequations} \begin{align}
  \Ih & = \int_0^\infty \Ih_\nu \, \dd\nuh, \\
  \jh & = \int_0^\infty \jh_\nu \, \dd\nuh, \\
  \alphahe & = \frac{1}{\Ih} \int_0^\infty \alphah_\nu \Ih_\nu \, \dd\nuh.
\end{align} \end{subequations}

The form of Equation~\eqref{eq:radiation_integral} informs a finite-volume/finite-solid-angle method, in which the conserved values $\nh^0 \nh_0 \Ih$ are updated not only via area-averaged spatial fluxes $\nh^i \nh_0 \Ih$, but also via angle-averaged angular fluxes $\nh^\zetah \nh_0 \Ih$ and $\nh^\psih \nh_0 \Ih$. The Ricci rotation coefficients hidden in $\nh^\zetah$ and $\nh^\psih$ account for the bending of light as it propagates on a curved spacetime, or for the rotation of the tetrad directions as one moves through space even in flat spacetime.

In a flat spacetime with $d$ spatial dimensions and a $(d{-}1)$-dimensional radiation angle space, one can always choose a tetrad such that $\omega^{\hat{\gamma}}_{\hat{\alpha}\hat{\beta}}$ vanishes everywhere, even when working in curvilinear coordinates with nonvanishing connection coefficients. Indeed, this has allowed the Newtonian predecessor of our algorithm to be used in 3D simulations in spherical coordinates with great success. However, given a symmetry of the spatial dimensions, there may be no such convenient tetrad that respects that symmetry. In particular, 2D axisymmetric simulations performed in cylindrical or spherical coordinates still require two angles to capture the radiation field, which is axisymmetric about a spatial axis but not any momentum axis, and so angular transport terms cannot be avoided even in such Newtonian problems. When general-relativistic gravity is present, angular transport will always be nontrivial for any choice of coordinates and tetrad, even for 3D simulations with two angular coordinates.

\subsection{Reconstruction}
\label{sec:method:reconstruction}

At the beginning of any stage in the time-integration algorithm, we have the primitive values $\Ih$ defined everywhere. This includes ghost zones in space and angle. We must calculate slope-limited ``left'' and ``right'' values at each spatial interface (at constant solid angle) and each angular interface (at constant spatial location).

Reconstruction in spatial directions is performed via the same one-dimensional algorithms as for cell-centered hydrodynamical variables, with each angular bin being treated separately. These routines include donor cell (DC), piecewise linear \citep[PLM,][]{VanLeer1974}, and piecewise parabolic \citep[PPM,][]{Colella1984} methods. Treating each angular bin separately is only valid to the extent that the tetrad basis vectors do not rotate too rapidly as one moves across the spatial grid. For example, for spatial cells indexed by $i$, $j$, and $k$ and angular bins indexed by $l$ and $m$, $n_{i,j,k,l,m}$ should point in roughly the same direction as $n_{i\pm1,j,k,l,m}$. This is similar to how, in hydrodynamics on a spherical grid, the azimuthal cells must not be so large that the $\phi$-direction changes wildly from one cell to the next. In practice, reasonable tetrad choices will result in valid radiation reconstruction on any grid with sufficient resolution for hydrodynamics.

Angular fluxes are computed similarly within each spatial cell. The procedure is most easily illustrated with the PLM stencil on a latitude-longitude grid. For concreteness, consider a grid with $N_\zetah = 4$, indexed by $l \in \{0,\, \ldots,\, 3\}$, and $N_\psih = 8$, indexed by $m \in \{0,\, \ldots,\, 7\}$, with half-integers denoting interfaces. Second-order reconstruction of left and right values of $\Ih$ at $(l, m) = (1, 1/2)$, with coordinates $(\zetah, \psih) = (3 \pi / 8, \pi / 4)$, would use angular cells with $(l, m) = (1, 7),\, (1, 0),\, (1, 1),\, (1, 2)$, with coordinates understood to be $\psih = -\pi / 8,\, +\pi / 8,\, +3 \pi / 8,\, +5 \pi / 8$. The left and right values at $l = 1/2$ and $m = 0$ would require crossing the pole, using the four cells with $(l, m) = (0, 4),\, (0, 0),\, (1, 0),\, (2, 0)$ and understood coordinates $\zetah = -\pi / 8,\, +\pi / 8,\, +3 \pi / 8,\, +5 \pi / 8$. An alternative method appropriate for geodesic grids is discussed in Section~\ref{sec:method:angular_grids}.

As $\Ih$ is essentially a scalar (it has no directional sense independent of the angular cell in which it is located), there is no sign change across the angular grid poles that must be considered as happens, for example, when reconstructing velocity vectors across the poles in hydrodynamics on spherical grids.

\subsection{Fluxes}
\label{sec:method:fluxes}

In vacuum, radiation fluxes are simply given by the upwind primitives multiplied by the appropriate prefactors, where the upwind direction is determined by the normal direction of the angular cell in question. That is, we take
\begin{equation} \label{eq:flux_space}
  \ave[\big]{\ncon{a} \ncov{0} \Ih}_{t A_a \Omegah} = \ncon{a} \ncov{0}
  \begin{cases}
    \Ihl, & \ncon{a} > 0, \\
    \Ihr, & \ncon{a} < 0,
  \end{cases}
\end{equation}
where $\Ihl$ and $\Ihr$ are the reconstructed intensities at the lesser and greater coordinate $x^a$ sides of the constant-$x^a$ interface. Angular fluxes are just as simple in our notation:
\begin{equation} \label{eq:flux_angle}
  \ave[\big]{\ncon{\lh} \ncov{0} \Ih}_{t V \lambda_\lh} = \ncon{\lh} \ncov{0}
  \begin{cases}
    \Ihl, & \ncon{\lh} > 0, \\
    \Ihr, & \ncon{\lh} < 0.
  \end{cases}
\end{equation}

In principle, a radiation--hydrodynamics scheme could simply use Equations~\eqref{eq:flux_space} and~\eqref{eq:flux_angle} for the radiation sector, with the hydrodynamics sector using its own Riemann-solver fluxes, which do not depend on the radiation. In practice, however, such operator splitting can lead to excessive diffusion when the speed of light greatly exceeds the effective diffusion of radiation through an optically thick medium.

Radiation--hydrodynamics codes employ various methods, similar in spirit but differing in detail, to overcome this issue. For example, \citet{Sadowski2013} uses a Riemann solver to calculate radiation fluxes, with the relevant signal speed estimates (nominally the eigenvalues of the linearized, homogeneous equations for the radiation subsector) are reduced toward $0$ in regions of large optical depth per cell. Similarly, \citet{Foucart2015} and \citet{Ryan2020} use the optical depth per cell to interpolate between a free-streaming flux and an advective--diffusive flux.

Our own approach most closely resembles that of \citet{Jiang2014}, generalized to a relativistic setting. For each interface, the primitive velocities $u^{a'}$ and fluid-frame total absorption coefficients $\alphabe$ are interpolated to the interface from the two adjacent cells. Given a timestep $\Delta t$, this defines an effective optical depth for the interface of
\begin{equation}
  \tau \propto \frac{\alphabe \Delta t}{u^\zeroh},
\end{equation}
where a proportionality constant of approximately $1$ to $10$ can be chosen based on the problem at hand, and a value of $0$ reduces to the free-streaming case. We next define an interpolating factor
\begin{equation}
  f = -\expm(-\tau^2),
\end{equation}
where $\expm(x) \equiv \exp(x) - 1$ is evaluated numerically without loss of precision near $x = 0$. Using the velocities from the adjacent cells, we can calculate $\nh^\zerob_\mathrm{L/R} = -(u_\alpha \nh^\alpha)_\mathrm{L/R}$ in those cells. The flux is then calculated according to
\begin{equation}
  \ave[\big]{\ncon{a} \ncov{0} \Ih}_{t A_a \Omegah} = \ncon{a} \ncov{0}
  \begin{cases}
    \Ihl (1 - g_\mathrm{L} + g), & \ncon{a} > 0, \\
    \Ihr (1 - g_\mathrm{R} + g), & \ncon{a} < 0,
  \end{cases}
\end{equation}
where we define
\begin{equation}
  g_\mathrm{L/R} = -f \paren[\bigg]{1 + \frac{1}{\nh^\zerob_\mathrm{L/R}}}
\end{equation}
and $g$ is the interpolation of $g_\mathrm{L/R}$ to the interface.

The discretization of space and momentum space naturally results in numerical diffusion. For example, beams of light will slowly spread beyond the region allowed by an exact solution to the geodesic equation (see Section~\ref{sec:tests:curvilinear_beams}). This can lead to a subtle numerical issue inside the ergosphere of a black hole, where negative-energy orbits exist.

The covariant momentum of radiation can be written
\begin{equation}
  \kcov{\alpha} = \nuh \ncov{\alpha}.
\end{equation}
Along a null geodesic in a stationary spacetime, $k_0$ is preserved. As $\nuh$ is strictly positive, the sign of $\nh_0$ must be preserved as well. However, numerical diffusion can send some light from a cell and angular bin with one sign of $\nh_0$ to a cell and angular bin with the other sign. If this is the case, and if the sign of $\nh_0$ on the interface, whether spatial or angular, matches the sign downwind, then any flux across the interface will increase $\Ih$ on both sides, leading to an exponential growth in $\Ih$.

This happens despite conservation of energy; the energy at infinity of the radiation field does not change, since light is being added to both positive-~and negative-energy orbits simultaneously. Increasing resolution can decrease the phase-space volume affected by this instability, but it will not decrease the number of cells on a particular grid where this will occur. To quell the instability, we introduce a critical value $\nzerocrit \approx 0.1$. Every step, we zero $\Ih$ wherever $\abs{\nh_0} < \nzerocrit$. This occurs in a relatively small fraction (less than $11\%$) of the phase-space volume between the horizon and the ergosphere.

\subsection{Coupling}
\label{sec:method:coupling}

We apply the effect of the right-hand side of Equation~\eqref{eq:radiation_integral} (or, more simply, the gray version of Equation~\eqref{eq:radiation_differential}, given that source terms affect each cell in isolation) each stage in the time integrator, after both the hydrodynamic and radiation quantities have been updated via flux divergences. Here we consider emission and true absorption (no stimulated emission) that are isotropic in the fluid frame. For concreteness, and to match the conditions of our initial applications, we specialize to photons (as opposed to neutrinos) interacting with a gamma-law gas (as opposed to one with a different equation of state). However, the procedure we outline is adaptable to these other cases.

Given the stationary nature of the tetrad, Equation~\eqref{eq:stationary_tetrad}, we can take the coupling portion of the equation, operator-split from the spatial and angular transport, to be
\begin{equation} \label{eq:rad_source}
  \pp{0} \Ih = \frac{1}{\nh^0} \paren[\big]{\jh - \alphahe \Ih}.
\end{equation}
Using the Lorentz-invariant forms of $I$, $j$, and $\alpha$, as well as the fact that the ratio of frequencies seen in different frames is that of the time components of $\nh$ (e.g., $\nub / \nuh = \nh^\zerob / \nh^\zeroh = \nh^\zerob$), the equation can be transformed to the fluid frame:
\begin{equation}
  \pp{0} \Ib = \frac{\nh^\zerob}{\nh^0} \paren[\big]{\jb - \alphabe \Ib}.
\end{equation}
Here, we implicitly assume the fluid velocity does not appreciably change over the course of the coupling, so that $\nh^\zerob$ has no time derivative.

Consider the decomposition of emissivity and absorptivity into absorption and scattering terms,
\begin{subequations} \begin{align}
  \jb & = \jba + \jbs, \\
  \alphabe & = \alphabea + \alphabes.
\end{align} \end{subequations}
Assuming the fluid remains in local thermodynamic equilibrium with temperature $T$, the gray form of Kirchhoff's law tells us the absorption terms are related via
\begin{equation}
  \jba = \frac{1}{4 \pi} \alphabpa \arad T^4,
\end{equation}
where
\begin{equation}
  \alphabpa = \frac{4 \pi}{\arad T^4} \int_0^\infty \alphaba_\nu B_\nu \, \dd\nu
\end{equation}
is the Planck-mean absorption coefficient, $B_\nu$ is the Planck function, and $\arad$ is the radiation constant. With scattering that is elastic,
\begin{equation}
  \oint \paren[\big]{\jbs - \alphabes \Ib} \, \dd\Omegab = 0,
\end{equation}
and fluid-frame isotropic, we have the relation
\begin{equation}
  \jbs = \frac{1}{4 \pi} \alphabes \Eb,
\end{equation}
where
\begin{equation}
  \Eb = \oint \Ib \, \dd\Omegab = \oint \Ih \ncon{\zerob} \ncon{\zerob} \, \dd\Omegah
\end{equation}
is the radiation energy density in the fluid frame. Combining these results, the coupling equation can be written
\begin{equation} \label{eq:explicit_coupling}
  \pp{0} \Ib = \frac{\nh^\zerob}{\nh^0} \paren[\bigg]{\frac{1}{4 \pi} \paren[\big]{\alphabpa \arad T^4 + \alphabes \Eb} - \alphabe \Ib}.
\end{equation}

In principle, one could apply Equation~\eqref{eq:explicit_coupling} cell-by-cell and angle-by-angle as an explicit update. However, the coupling coefficients $\alphabpa$, $\alphabes$, and/or $\alphabe$ might be large enough that their associated timescales (their reciprocals, in units where $c = 1$) are much shorter than the hydrodynamic $\Delta t$. Accuracy and stability demand an implicit solution to the equation. Fortunately, a locally implicit solution suffices in relativity, as the hydrodynamic evolution is already limited by light-crossing times. Naively, the appearance of $\Eb$ in Equation~\eqref{eq:explicit_coupling} would couple all angles in a cell together, resulting in an implicit system whose size scales with the number of angles used. However, as shown in \citet{Jiang2021} for the Newtonian case, under the small-velocity-shift assumption already made, the solution can be obtained by transforming intensities into the fluid frame and solving a quartic equation for the post-coupling temperature. This method has been extended to SR by \citet{Zhang2021}, and here we extend it to the fully general-relativistic case.

Let $-$ and $+$ indicate quantities before and after coupling. Implicit differencing of Equation~\eqref{eq:explicit_coupling} yields
\begin{multline} \label{eq:implicit_ii}
  \Ib_+ - \frac{\nh^0}{\nh^0 + \nh^\zerob \alphabe \Delta t} \Ib_- \\ - \frac{\nh^\zerob \Delta t}{4 \pi (\nh^0 + \nh^\zerob \alphabe \Delta t)} \paren[\big]{\alphabpa \arad T_+^4 + \alphabes \Eb_+} = 0.
\end{multline}
Integrating over solid angle in the fluid frame, we have the relation
\begin{equation} \label{eq:coupling_rad}
  \Eb_+ - A_1 - A_2 \paren[\big]{\alphabpa \arad T_+^4 + \alphabes \Eb_+}  = 0,
\end{equation}
where we define the coefficients
\begin{subequations} \begin{align}
  A_1 & = \oint \frac{\nh^0}{\nh^0 + \nh^\zerob \alphabe \Delta t} \Ib_- \, \dd\Omegab, \\
  A_2 & = \frac{\Delta t}{4 \pi} \oint \frac{\nh^\zerob}{\nh^0 + \nh^\zerob \alphabe \Delta t} \, \dd\Omegab,
\end{align} \end{subequations}
and where we again assume the fluid velocity changes negligibly over the timestep in order to write
\begin{equation}
  \Eb_+ = \oint \Ib_+ \, \dd\Omegab_-.
\end{equation}

At the same time, to the extent the fluid frame is not accelerated, total energy conservation relates the change in fluid internal energy to the change in fluid-frame radiation energy. More precisely, with the assumptions of a stationary tetrad and fluid velocity, we can multiply Equation~\eqref{eq:rad_source} by $\nh^0 \nh_\alpha$ and integrate over tetrad-frame solid angle to find
\begin{equation}
  \pp{0} \Rsplit{0}{\alpha} = -\Gcov{\alpha}.
\end{equation}
Here we use the radiation stress--energy tensor, whose components\footnote{This equation can be written in a variety of equivalent forms, with the four free indices agreeing in coordinate system and covariance/contravariance, and with the four macrons agreeing (e.g., all changed to bars or all eliminated).} are
\begin{equation}
  \Rcon{\alpha\beta} = \oint \Ih \ncon{\alpha} \ncon{\beta} \, \dd\Omegah.
\end{equation}

Combined with the operator-split portion of Equation~\eqref{eq:grmhd_differential:momentum},
\begin{equation}
  \pp{0} \Tsplit{0}{\alpha} = \Gcov{\alpha},
\end{equation}
the stationary assumption allows us to write
\begin{equation}
  \pp{0} \paren[\big]{\Rcon{\zerob\alphab} + \Tcon{\zerob\alphab}} = 0.
\end{equation}
As we have $R^{\zerob\zerob} = \Eb$ and $T^{\zerob\zerob} = \rho + \ugas + \umag$, with $\ugas$ and $\umag$ the thermal and fluid-frame electromagnetic energy densities, and knowing radiation does not couple directly to magnetic fields, this allows us to write
\begin{equation} \label{eq:coupling_tot}
  \frac{\kB \rho}{\mu \mprot (\Gamma - 1)} (T_+ - T_-) = -\paren[\big]{\Eb_+ - \Eb_-},
\end{equation}
where we use the ideal gas equation of state and leverage the fact that this coupling cannot change density.\footnote{More accurately, the conserved density $\rho u^0$ is unchanged by coupling (cf.\ Equation~\eqref{eq:grmhd_differential:momentum}). With the assumption of a small shift in velocity made twice already, this amounts to $\rho$ being constant, which is needed for Equation~\eqref{eq:coupling_tot} to hold.}

Combining Equations~\eqref{eq:coupling_rad} and~\eqref{eq:coupling_tot}, we have a quartic equation for $T_+$:
\begin{equation}
  B_4 T_+^4 + B_1 T_+ + B_0 = 0,
\end{equation}
with
\begin{subequations} \begin{align}
  B_4 & = A_2 \alphabpa \arad, \\
  B_1 & = \frac{\kB \rho}{\mu \mprot (\Gamma - 1)} (1 - A_2 \alphabes), \\
  B_0 & = A_1 - B_1 T_- - (1 - A_2 \alphabes) \Eb_-.
\end{align} \end{subequations}
With $T_+$ in hand, Equation~\eqref{eq:coupling_tot} yields $\Eb_+$. The new fluid-frame intensities can be found from Equation~\eqref{eq:implicit_ii}, after which they can be transformed to the tetrad frame.

When the dynamics are dominated by radiation, which we take to be when $\Eb$ exceeds the gas inertial density $\rho + \ugas$, the assumption of fluid velocity not changing over the course of coupling no longer holds. In such cases we modify the above procedure by using a different fluid velocity, taken to be the same before and after coupling, that approximates the velocity we expect the gas to have as a result of coupling.

Given the radiation stress--energy components $R^{ab}$, we define a ``radiation three-velocity'' via
\begin{equation}
  \vradcon{\ah} = \frac{R^{\zeroh\ah}}{R^{\zeroh\zeroh}},
\end{equation}
together with its associated four-velocity. This allows us to calculate the momentum density of the gas both at its pre-coupling velocity and at the radiation velocity:
\begin{subequations} \begin{align}
  \Tgascon{\zeroh\ah} & = (\rho + u + p) \ugascon{\zeroh} \ugascon{\ah}, \\
  \Tradcon{\zeroh\ah} & = (\rho + u + p) \uradcon{\zeroh} \uradcon{\ah}.
\end{align} \end{subequations}
In the tetrad frame, the coupling force, Equation~\eqref{eq:coupling_force}, is simply
\begin{equation}
  \Gcon{\alphah} = -\paren[\big]{\alphabpa \arad T^4 + \alphabes \Eb} \ucon{\alphah} - \alphabe \ucov{\betah} \Rcon{\alphah\betah}.
\end{equation}
We use the fractions
\begin{equation}
  f^a = \frac{G^\ah \Delta t}{u^\ah (T_\mathrm{rad}^{\zeroh\ah} - T_\mathrm{gas}^{\zeroh\ah})},
\end{equation}
clamped between $0$ and $1$, to estimate
\begin{equation}
  \ucon{\ah} = (1 - f^a) \ugascon{\ah} + f^a \uradcon{\ah}
\end{equation}
for use in the implicit coupling procedure above.

After $I_+$ is found, the total four-momentum of the fluid and radiation is conserved by subtracting the change in ${R^0}_\alpha$ from ${T^0}_\alpha$.

In order to model neutrinos or non-gamma-law fluids, the details of the coupling procedure would change. In particular, the implicit solution need not reduce to solving a quartic equation. Still, it would be reducible to finding the root of a small number of equations.

\subsection{Angular Grids}
\label{sec:method:angular_grids}

Just as the use of spatial coordinates $x^a$ naturally leads to the discretization space via surfaces of constant $x^a$, our use of angular coordinates $\zetah$ and $\psih$ inform a discretization of angular space (the sphere) via lines of constant latitude and longitude. Our preceding discussion is framed in this way, and the code supports this style of angular grid.

By choosing $N_\zetah + 1$ lines of latitude equally spaced in $\cos\zetah$ from $+1$ to $-1$ and $N_\psih + 1$ lines of longitude equally spaced in $\psih$ from $0$ to $2 \pi$, we obtain $\Nang = N_\zetah N_\psih$ angular cells with equal solid angle, distributed roughly uniformly in the tetrad frame. The latitude/longitude grid scheme is conceptually simple, and it can flexibly support many reasonable values of $\Nang$ for different problems. However, such grids are not particularly isotropic:\ for $N_\psih = 2 N_\zetah$, the spherical triangles touching the pole have a short-to-long leg length ratio that scales as $\pi / N_\zetah$ for large $N_\zetah$.

\begin{figure}
  \centering
  \includegraphics{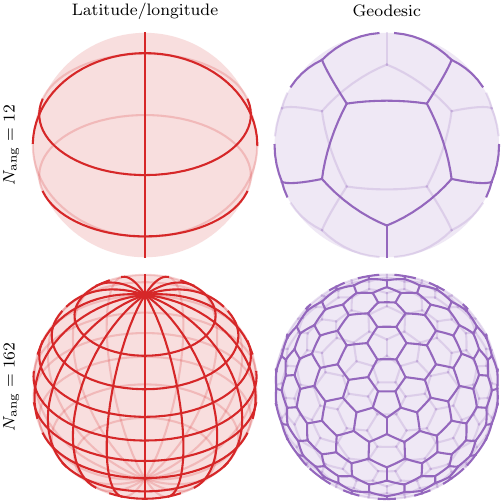}
  \caption{Examples of angular grids supported by the code. On the left are latitude/longitude grids with $3 \times 4$ angles (coarse, top) and $9 \times 18$ angles (fine, bottom). To the right of each is the geodesic grid with exactly the same number of cells (grid levels $1$ and $4$). The north pole (centered on a pentagonal cell in the geodesic grids) is tilted $35^\circ$ toward the camera in all cases. \label{fig:angular_grids}}
\end{figure}

We therefore implement a geodesic grid scheme as an alternative to latitude/longitude grids. A number of such schemes exist in the literature; here we choose the nonrecursive grid of \citet{Wang2011}, constructed conceptually as follows:
\begin{enumerate}
  \item Set $12$ points on the sphere to be the vertices of a regular icosahedron.
  \item Triangulate each of the $20$ faces of the icosahedron into $\Nlev^2$ equilateral triangles for a level $\Nlev$ grid, $\Nlev > 0$.
  \item Project the centers of all $20 \Nlev^2$ triangles onto the sphere, and connect via a spherical geodesic edge any pair of points originating from triangles sharing a common edge.
\end{enumerate}
The resulting discretization of the sphere has $\Nang = 10 \Nlev^2 + 2$ cells, $12$ of which are pentagons with the rest hexagons. This grid's anisotropy remains bounded as resolution increases, as shown by \citet{Wang2011}, and it has been successfully used for other astrophysical purposes \citep[e.g.,][]{Daszuta2021}. Figure~\ref{fig:angular_grids} illustrates coarse and fine versions of both the latitude/longitude and geodesic grids.

In hydrodynamics, grids with boundaries not coinciding with constant-coordinate surfaces present issues when solving Riemann problems. Vector quantities in the left and right states must be rotated into coordinates aligned with the grid, and the resulting fluxes will therefore also be aligned with the grid. However, fluxes of vector conserved quantities are components of a rank-$2$ tensor, and one-dimensional Riemann solvers do not provide enough components for this tensor to simply be rotated back into the coordinate frame in order to apply the flux-divergence update. With radiation this is not a problem, as our conserved quantity is a scalar related to intensity. That is, there is no rotation to perform on the reconstructed intensities, Equation~\eqref{eq:flux_angle} holds if the direction vector $\nh^\lh$ is understood to be the combination of $\nh^\zetah$ and $\nh^\psih$ aligned with the grid, and the angular flux-divergence term in Equation~\eqref{eq:radiation_integral} is naturally modified to sum over the (possibly more than four) edges of the cell.

Spatial fluxes of radiation, as in Equation~\eqref{eq:flux_space}, are not concerned with the discretization in angle. For the purposes of taking moments of the radiation field or coupling to matter, it is sufficient (i.e., second-order accurate) to know the angular grid's cell centers and solid angles. Reconstruction on latitude/longitude grids uses the same algorithms as spatial reconstruction. Geodesic grids can use DC reconstruction. For a higher-order technique, we use the minimum-angle plane reconstruction (MAPR) method developed for triangular meshes by \cite{Christov2008} and implemented on the same hexagonal grid as ours by \citet{Florinski2013}.

The geodesic grid sacrifices fine-grained control of resolution (supporting $\Nang \in \{12,\, 42,\, 92,\, 162,\, \ldots\}$), achieving instead a much more isotropic distribution of angles. In practice, this allows us to run simulations without being concerned that the grid might imprint numerical anisotropies on the solution. The bookkeeping of the geodesic grid is more complicated as well, but it can still lend itself to efficient memory packing and even vectorization by dividing it into five congruent, logically rectangular patches, as in \citet{Randall2002}.

Finally, for any type of grid with roughly equal solid angles, we can estimate the resolution needed to retain accuracy in the presence of strong Lorentz boosts. As a point is boosted by a three-velocity $v$ in a given direction, its sphere of angles distorts. Solid angles at the back of the sphere (i.e., in the direction opposite the boost) grow the most, changing by a factor approaching $(1 + v) / (1 - v)$ in the most anti-aligned case. If a problem requires that no angular bin in the fluid frame have a size greater than $\Delta\Omega_\mathrm{max}$, and if $v_\mathrm{max}$ is the largest fluid velocity relative to the tetrad frame that will arise, then the tetrad-frame angular grid should have
\begin{equation}
  \Nang \geq \frac{4 \pi}{\Delta\Omega_\mathrm{max}} \cdot \frac{1 + v_\mathrm{max}}{1 - v_\mathrm{max}}.
\end{equation}

\section{Tests}
\label{sec:tests}

\subsection{Colliding Beams}
\label{sec:tests:colliding_beams}

One of the primary motivations for the development of this method is the ability to capture radiation fields that are peaked in multiple directions simultaneously. Moment methods such as M1 fundamentally cannot do this, as is readily apparent in a colliding beam test. Two beams crossing in vacuum will merge into a single beam pointing in the average direction of the two when using M1. By discretizing in angle, our method trivially captures multimodal intensity distributions at each point.

On a $96 \times 60$ spatial grid covering $[0,\, 1.6] \times [0,\, 1]$, we place beam sources at $x = 2/15$ and $y = 1/2 \pm 11/30$, each with a radius of $1/10$. At each location in the source, radiation is emitted in a beam at a central angle $\mp \pi / 6$ with respect to the $+x$-axis, and the beams have a nonzero full spread of $\pi / 12$. Radiation is emitted into each appropriate angle at a rate of $\dd I / \dd t = 5$. The sources are optically thin:\ radiation emitted in one source cell is augmented rather than replaced by radiation in source cells through which it subsequently passes.

\begin{figure}
  \centering
  \includegraphics{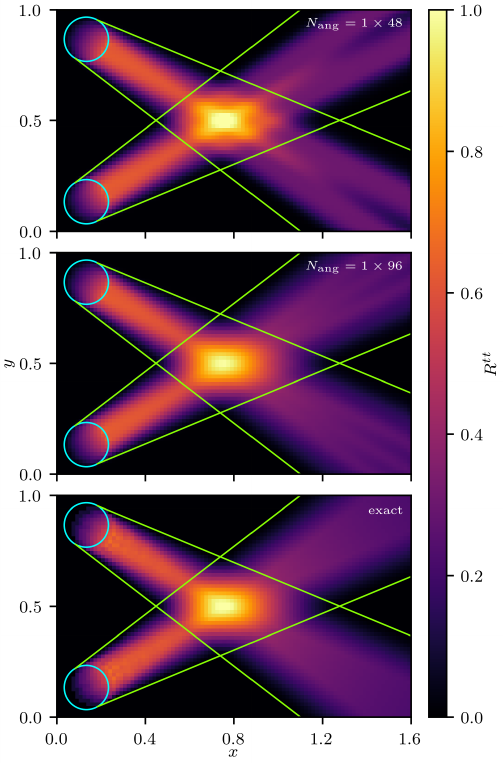}
  \caption{Solutions for the colliding beam test at $t = 5/2$ with two different resolutions, along with the exact solution. Both tests use $96 \times 60$ cells in space. Source regions are outlined in cyan. Green lines denote the maximum extent of radiation expected (i.e., the edge of the beam emitted from the corresponding edge of the source region). The beams easily cross without affecting one another at any resolution. \label{fig:colliding_beams}}
\end{figure}

Figure~\ref{fig:colliding_beams} illustrates the beam test with our code, using two different angular resolutions. The results are shown at $t = 5/2$, sufficient for the beams to leave the grid and for the problem to be in steady state. These beams are able to cross without affecting one another. Ray effects are more prominent at the lower resolution, where the full width of the beam is only twice the $\psih$-width of our cells in angle. For comparison, the exact solution is shown in the third panel. While simple, we emphasize that this test is failed by M1 and commonly employed closure methods.

\subsection{Beams in Curvilinear Coordinates}
\label{sec:tests:curvilinear_beams}

Any method that is to work with curved spacetimes should work in flat spacetime parameterized by non-Cartesian coordinates, where the connection coefficients $\Gamma^\alpha_{\beta\gamma}$ do not identically vanish. Even in Cartesian coordinates, a choice of tetrad might have nontrivial coordinate derivatives, leading to nonzero values for the Ricci rotation coefficients defined in Equation~\eqref{eq:ricci}. Here we illustrate the same physical problem in different flat-spacetime coordinate systems, each with different tetrads.

\begin{figure}
  \centering
  \includegraphics{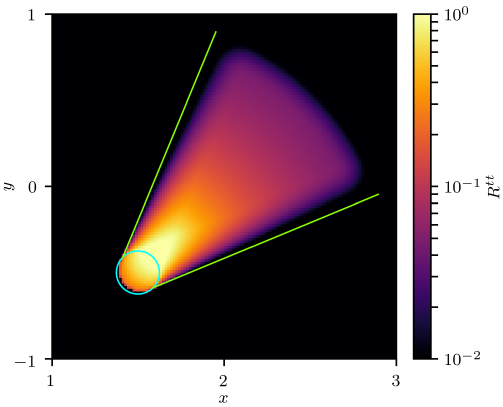}
  \caption{Exact midplane energy density expected for the curvilinear beam test in Figure~\ref{fig:curvilinear_beams}. As in Figure~\ref{fig:colliding_beams}, cyan lines enclose the source region while green lines indicate the maximum extent of radiation expected, including in distance from the source. \label{fig:curvilinear_exact}}
\end{figure}

Let the region within $1/8$ of the point $(x,\, y,\, z) = (3/2,\, -1/2,\, 0)$ be an optically thin source initially surrounded by no radiation. Let every point in the source emit radiation at a rate of $\dd I / \dd t = 10$ into any angle within $\pi / 8$ of the direction $(+1,\, +1,\, 0)$, and allow the simulation to run for a time of $\sqrt{2}$. The exact solution for this problem is shown in Figure~\ref{fig:curvilinear_exact}.

We perform this simulation with $64^3$ uniformly-spaced cells in Cartesian coordinates $(x,\, y,\, z) \in [1,\, 3] \times [-1,\, 1]^2$, cylindrical coordinates $(R,\, \phi,\, z) \in [1,\, \sqrt{10}] \times [-\pi / 4,\, \pi / 4] \times [-1,\, 1]$, and spherical coordinates $(r,\, \theta,\, \phi) \in [1,\, \sqrt{10}] \times [\pi / 4,\, 3 \pi / 4] \times [-\pi / 4,\, \pi / 4]$. We additionally add a single level of static mesh refinement to a quarter of the domain ($x < 2$, $y < 0$; $R < (1 + \sqrt{10}) / 2$, $\phi < 0$; or $r < (1 + \sqrt{10})$, $\phi < 0$) encompassing the source, demonstrating the ability of our code to transport radiation through mesh refinement boundaries.

\begin{figure*}
  \centering
  \includegraphics{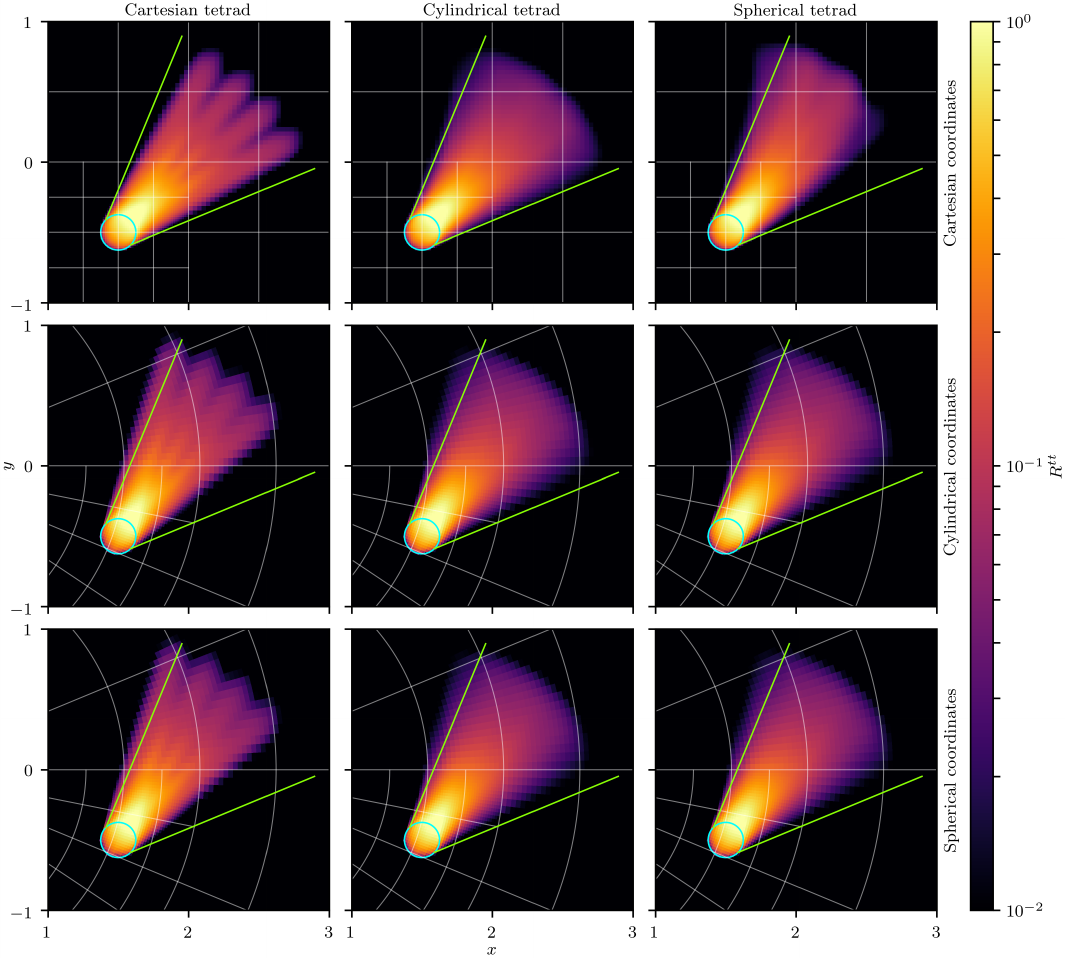}
  \caption{Midplane energy densities for the beam test in three coordinate systems and three choices of tetrad. All tests use $64^3$ cells in space and $16 \times 32$ angles. Blocks of $16^2$ cells are outlined by the grid. In all cases the numerics match the expected result shown in Figure~\ref{fig:curvilinear_exact}, though beam effects and anisotropic numerical diffusion can be seen with the logarithmic color scale. \label{fig:curvilinear_beams}}
\end{figure*}

In each coordinate system we use either a Cartesian tetrad with $n^3$ (the north pole of the angular grid) in the $z$-direction and $n^1$ (the $0$-longitude point on the equator) in the $x$-direction, a cylindrical tetrad with $n^3$ in the $z$-direction and $n^1$ in the $R$-direction, or a spherical tetrad with $n^3$ in the $\theta$-direction and $n^1$ in the $\phi$-direction. In all cases we use $16 \times 32$ angles. The resulting energy densities in the $z = 0$ planes of the nine simulations are shown in Figure~\ref{fig:curvilinear_beams}.

In all cases, the beam travels at the speed of light in the correct direction with the correct spread, with no artifacts at the refinement boundary. There are angular transport terms when the cylindrical and spherical tetrads are used. These result in some anisotropic numerical diffusion of the radiation, seen by the beam energy not being centered between the two extremal geodesics, but this is much less perceptible when using a color scale that is linear in energy. This diffusion has the property of removing ray effects, from which one can clearly see that only four angular bins in azimuth are populated in the cases with the Cartesian tetrad.

In order to illustrate our method's capability in coordinate systems with non-diagonal metrics, we consider snake coordinates as in \citet{White2016}. That is, in terms of Minkowski $(t,\, x,\, y,\, z)$, we have snake $(t,\, w,\, y,\, z)$, with
\begin{equation}
  w = x - A \sin(k \pi y).
\end{equation}
The metric becomes\footnote{The notation here differs slightly from \citet{White2016}, and we correct a typographical error in that work.}
\begin{equation}
  \gcov{\alpha\beta} =
  \begin{pmatrix}
    -1 & 0 & 0 & 0 \\
    0 & 1 & \delta & 0 \\
    0 & \delta & 1 + \delta^2 & 0 \\
    0 & 0 & 0 & 1
  \end{pmatrix},
\end{equation}
with
\begin{equation}
  \delta = A k \pi \cos(k \pi y).
\end{equation}
In this test we set $A = 0.1$ and $k = 2$.

We run this problem on a snake grid with $128^2$ cells covering $[-0.5,\, 0.5]$ in $w$ and $[-0.05,\, 0.95]$ in $y$. Here we showcase a level-$22$ geodesic grid ($\Nang = 4842$). The beam is centered at the origin with a proper diameter of $0.05$. It points vertically in Minkowski coordinates with a full spread of $10^\circ$. The problem is run to a time of $t = 3$, at which point the leading edge of the beam has long since left the grid.

\begin{figure*}
  \centering
  \includegraphics{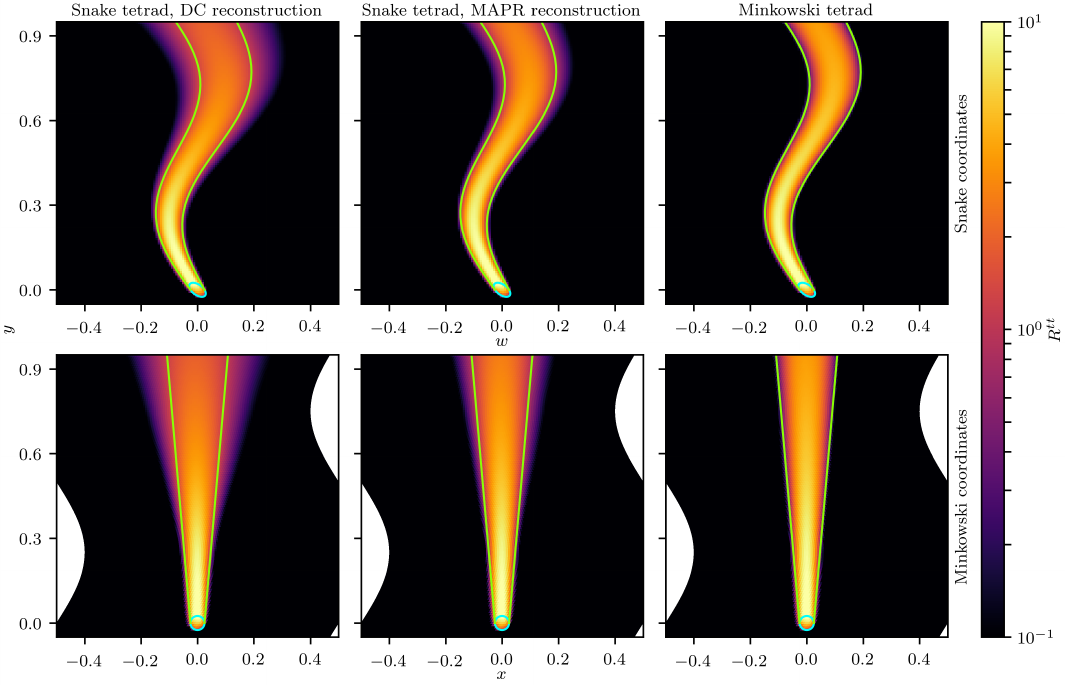}
  \caption{Energy densities for the beam test in Minkowski spacetime with snake coordinates. The grid has $128^2$ cells and $4842$ angles. Aligning the tetrad with snake coordinates results in numerical diffusion governed by the angular reconstruction algorithm, while aligning the tetrad with Minkowski coordinates eliminates this effect. The lower panels show the solution transformed back into Minkowski coordinates, where the beam can be seen to always propagate in the correct direction. \label{fig:snake_beams}}
\end{figure*}

Figure~\ref{fig:snake_beams} shows the results of this test. In two cases we use a snake-coordinate-aligned tetrad. As geodesics are trivial in Minkowski but not snake coordinates, aligning the tetrad in this way necessitates nonzero angular fluxes stemming from nonvanishing Ricci coefficients. The left panels illustrate this case with DC reconstruction, which results in a substantial amount of numerical diffusion. In the middle, we use MAPR reconstruction (see Section~\ref{sec:method:angular_grids}), which reduces diffusion. On the right we show what happens when the tetrad is aligned with Minkowski directions rather than snake directions:\ angular fluxes and numerical diffusion vanish, despite the beam not traveling along a coordinate direction (that is, despite the connection coefficients not all vanishing). The upper panels show the results in the same coordinates in which the test is run; the lower panels transform the outputs to Minkowski coordinates.

These tests demonstrate how numerical artifacts can vary with choice of tetrad, just as they can vary with choice of coordinates. Our method grants users significant freedom in choosing tetrads appropriate for the problem at hand.

\subsection{Beams around Black Holes}
\label{sec:tests:black_hole_beams}

We proceed with beam tests in a fully general-relativistic setting, in the vicinity of a black hole.

\begin{figure}
  \centering
  \includegraphics{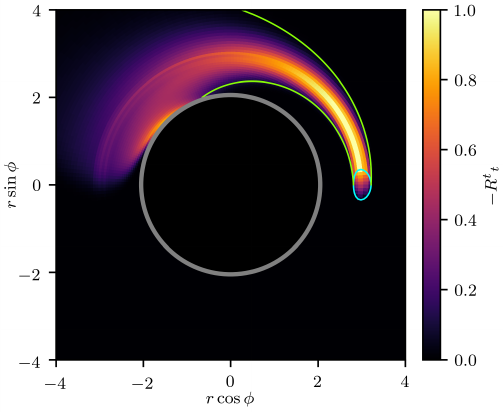}
  \caption{Equatorial beam test around a nonspinning black hole. Here we have $54 \times 5 \times 256$ cells (out to $r = 6$) and $15 \times 128$ angles. The beam correctly propagates halfway around the photon sphere. The gray region denotes locations outside the horizon but off the grid, while the black hole itself is delineated with the inner black disk. The increase in energy near the inner boundary is an expected feature of non-horizon-penetrating coordinates. \label{fig:schwarzschild_equatorial_beam}}
\end{figure}

First, we consider a nonspinning black hole in Schwarzschild coordinates $(t,\, r,\, \theta,\, \phi)$. We construct a spatial grid on $[2.1,\, 6] \times [123 \pi / 256,\, 133 \pi / 256] \times [0,\, 2 \pi]$ using $54 \times 5 \times 256$ cells, logarithmically spaced in $r$ and uniformly spaced in angle. This grid focuses on the vicinity of the midplane, avoids the singularity at the horizon, and keeps the proper aspect ratios of the cells approximately unity. We employ a cylindrical tetrad, with $\nh^\threeh$ and the north pole of the angular grid pointing in the direction of increasing $z = r \cos\theta$, and with $\nh^\oneh$ pointing in the direction of increasing $R = r \sin\theta$. For this test we use $15 \times 128$ angles. The source region is defined to be everywhere within a proper distance of $7/20$ of $(3,\, \pi / 2,\, 0)$, with the beam pointing in the prograde azimuthal direction with a half-opening angle of $5^\circ$. Radiation is injected at a rate of $\dd\Ih / \dd t = 30$. The resulting conserved energy density at time $t = 3 \sqrt{3} \pi$ is shown in Figure~\ref{fig:schwarzschild_equatorial_beam}.

The beam correctly moves along the photon sphere at $r = 3$, spreading as expected given the nonzero source size and opening angle. The duration of the simulation is chosen to correspond to half an orbit along the photon sphere, and indeed the middle of the beam travels $180^\circ$ around the black hole. Note that the conserved energy $-{R^t}_t$ increases toward the inner boundary of the simulation. This results from using non-horizon-penetrating coordinates, in which radiation energy can only asymptotically approach the horizon but never pass through it, causing the energy to grow with decreasing radius in steady state.

\begin{figure}
  \centering
  \includegraphics{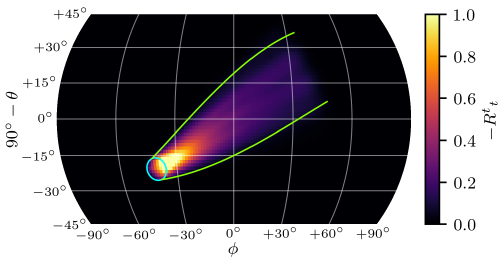}
  \caption{Inclined beam test around a nonspinning black hole. Color indicates the conserved energy density of the $r = 3$ slice, shown via a Mollweide projection. We use $5 \times 64 \times 128$ cells and $16 \times 32$ angles. The beam correctly propagates a quarter orbit around the photon sphere, making an angle of $30^\circ$ with the spatial grid as it crosses the midplane. \label{fig:schwarzschild_inclined_beam}}
\end{figure}

Next, we launch a beam at an angle to the grid, still along the photon sphere. For this test, we use $64 \times 128$ cells in angular coordinates, covering $[\pi / 4,\, 3 \pi / 4] \times [-\pi / 2,\, +\pi / 2]$. In the radial direction, we have $5$ cells logarithmically spaced from approximately $2.896$ to $3.108$, again resulting in equal aspect ratios. We use a spherical tetrad ($\nh^\threeh$ in the $\theta$-direction, $\nh^\oneh$ in the $\phi$-direction) with $16 \times 32$ angles. The source and beam centers are set such that after one-eighth of an orbit the beam will cross the midplane at a $30^\circ$ angle. The source has a proper radius of $1/4$, and the beam has a half-opening angle of $15^\circ$. Here we use $\dd\Ih / \dd t = 8$. The result at $t = 3 \sqrt{3} \pi / 2$ (a quarter orbit) is shown in Figure~\ref{fig:schwarzschild_inclined_beam}.

This beam also propagates in the correct direction at the correct speed. We note that the beam does not align with the $\theta$-~or $\phi$-directions, nor does it align with the $\zetah$-~or $\psih$-directions in angular space.

\begin{figure}
  \centering
  \includegraphics{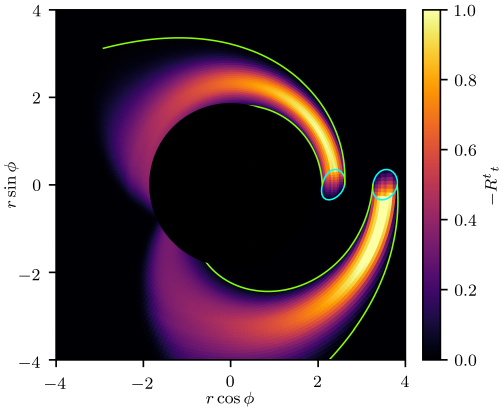}
  \caption{Prograde and retrograde equatorial beam tests around a spinning black hole with $a = 1/2$. The grid employs $62 \times 5 \times 256$ cells (from inside the horizon out to $r = 6$) and $15 \times 128$ angles. The midplane conserved energy density shows the beams propagating the correct distances of half an orbit and approximately one-third of an orbit. The region inside the horizon is masked in black. \label{fig:kerr_beams}}
\end{figure}

For a third test of beams in GR, we consider a black hole with dimensionless spin $a = 1/2$, described in spherical Kerr--Schild coordinates $(t,\, r,\, \theta,\, \phi)$. With the introduction of spin, the prograde and retrograde photon orbits separate to radii
\begin{equation}
  r_\pm = 2 + 2 \cos \paren[\bigg]{\frac{2}{3} \cos^{-1} (\mp a)}.
\end{equation}
We place a source (proper diameter $7/20$) at both radii, each with an azimuthal beam pointed in the appropriate direction (half-opening angle $5^\circ$), as shown in Figure~\ref{fig:kerr_beams}. Radiation is injected at the prograde and retrograde sources at rates of $\dd\Ih / \dd t = 70$ and $\dd\Ih / \dd t = 60$, respectively.

For this test, we use a $62 \times 5 \times 256$ grid in space, covering approximately $[1.831,\, 6] \times [1.508,\, 1.634] \times [0,\, 2 \pi]$, with logarithmic spacing in radius. This grid again makes the proper aspect ratios unity:\ $\sqrt{g_{rr}} \Delta r = \sqrt{g_{\theta\theta}} \Delta\theta = \sqrt{g_{\phi\phi}} \Delta\phi$, calculated at the midplane and at the geometric mean of $r_+$ and $r_-$. As Kerr--Schild coordinates are horizon penetrating, we place a single cell inside the horizon. We use a spherical tetrad with $15 \times 128$ angles. At time $t = \pi (r_+^{3/2} + a)$, the prograde beam should propagate halfway around the black hole, while the retrograde beam should complete a fraction $(1/2) (r_+^{3/2} + a) / (r_-^{3/2} - a) \approx 1/3$ of its orbit. Indeed, this is exactly what we see.

\subsection{Hohlraums}
\label{sec:tests:hohlraum}

Following \citet{Ryan2020}, we consider an infinite wall and the 1D problem of the propagation of radiation off the wall into vacuum. The wall maintains a fixed, isotropic radiation field with energy density $1$. After a time $t$ and at a distance $x < t$ from the wall, the radiation field should have intensity $1/4 \pi$ for all angles with $n^x > x / t$. Thus we have the analytic solution for the moments
\begin{subequations} \begin{align}
  \Rcone{tt} & = \frac{1}{2} \paren[\bigg]{1 - \frac{x}{t}}, \\
  \Rcone{tx} & = \frac{1}{4} \paren[\bigg]{1 - \frac{x^2}{t^2}}, \\
  \Rcone{xx} & = \frac{1}{6} \paren[\bigg]{1 - \frac{x^3}{t^3}},
\end{align} \end{subequations}
with all moments vanishing for $x > t$.

\begin{figure}
  \centering
  \includegraphics{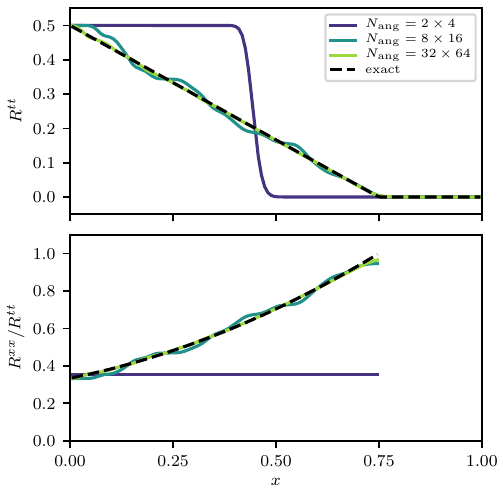}
  \caption{Solutions for the 1D, plane-parallel hohlraum test with $128$ cells in $x$ using latitude/longitude grids. Eddington factors are only shown where they are well defined in the exact solution. With enough angles, we accurately capture the radiation field. \label{fig:hohlraum_1d_solution}}
\end{figure}

\begin{figure}
  \centering
  \includegraphics{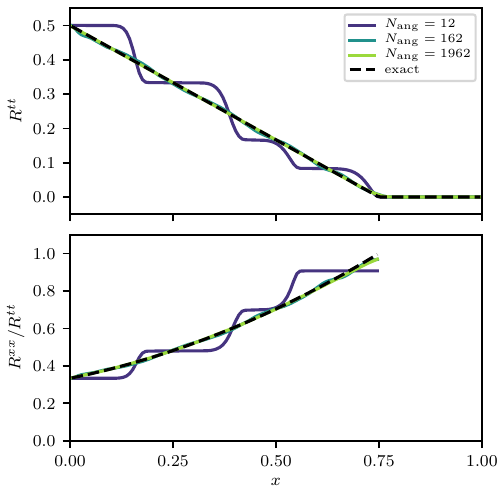}
  \caption{Solutions for the 1D, plane-parallel hohlraum test with $128$ cells in $x$ using geodesic grids. The resolutions here are comparable to those in Figure~\ref{fig:hohlraum_1d_solution}. \label{fig:hohlraum_1d_solution_geodesic}}
\end{figure}

We use $128$ cells in $x$, and consider latitude/longitude grids with $N_\zetah$ ranging from $2$ to $32$. We keep $N_\psih = 2 N_\zetah$. Figure~\ref{fig:hohlraum_1d_solution} shows the energy density and Eddington factor at $t = 3/4$ for three different angular resolutions; Figure~\ref{fig:hohlraum_1d_solution_geodesic} shows the results obtained with three geodesic grids with comparable resolutions. While $8\text{--}12$ angles fail to capture nuances in the radiation field, higher resolutions rapidly approach the analytic solution.

\begin{figure}
  \centering
  \includegraphics{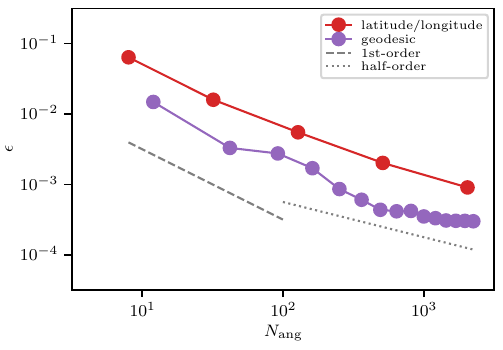}
  \caption{Convergence of the 1D, plane-parallel hohlraum test with angular resolution. For all tests we use $128$ cells in $x$. Half-order convergence ($\epsilon \sim \Nang^{-1/2}$) is expected for a test with a discontinuous radiation field. Eventually, errors due to the fixed spatial grid in this test dominate and convergence stops. \label{fig:hohlraum_1d_convergence}}
\end{figure}

We calculate an error according to
\begin{subequations} \begin{align}
  \epsilon & = \frac{1}{\sqrt{3}} \parenpow{\big}{\epsilon_{tt}^2 + \epsilon_{tx}^2 + \epsilon_{xx}^2}{1/2}, \\
  \epsilon_{\alpha\beta} & = \int_0^1 \abs[\big]{\Rcon{\alpha\beta} - \Rcone{\alpha\beta}} \, \dd x.
\end{align} \end{subequations}
The results are shown in Figure~\ref{fig:hohlraum_1d_convergence} for five latitude/longitude grids and for all geodesic grids up to $\Nlev = 15$ ($\Nang = 2252$). We expect to see $\epsilon \sim \Nang^{-1/2}$, given the discontinuous nature of the radiation field in angle at all times and locations, together with the fact that each of our angular bins in this problem is essentially ``on'' or ``off.'' We see even faster convergence at low angular resolutions. Eventually, the errors in the geodesic grid reach a noise floor, which is expected given the finite spatial resolution. For comparison, the MOCMC method \citep{Ryan2020} achieves convergence with samples as $\epsilon \sim \Nsamp^{-1}$ by using the sample points to perform high-order quadrature on the sphere. As $\Nsamp$ ranges from $16$ to $512$, the amplitude of the MOCMC errors is comparable to what we find with the latitude/longitude grid over the same range of $\Nang$.

\Citet{Ryan2020} also consider a 2D version of the hohlraum test, where radiation from a left and a bottom wall propagates into vacuum. The other two walls employ reflecting boundary conditions; equivalently, the radiating walls extend far enough for their ends to not be in causal contact with the solution of interest.

This problem also has a closed-form solution. After time $t$ and at a distance $x < t$ from the left wall and $y$ from the bottom wall, the energy density seen from the left wall is
\begin{equation}
  \Rconex{tt} = \frac{1}{2} - \frac{(\pi - \eta) x}{2 \pi t} - \frac{1}{2 \pi} \sin^{-1} \paren[\Bigg]{\frac{x \sin\eta}{\sqrt{x^2 + y^2}}},
\end{equation}
where we define the first-quadrant angle $\eta$ via
\begin{equation}
  \cos\eta = \min \paren[\bigg]{\frac{y}{\sqrt{t^2 - x^2}}, 1}.
\end{equation}
Interchanging $x$ and $y$ yields the energy density $R_{\mathrm{exact},y}^{tt}$ from the bottom wall. The total energy density is the sum of the two.

\begin{figure}
  \centering
  \includegraphics{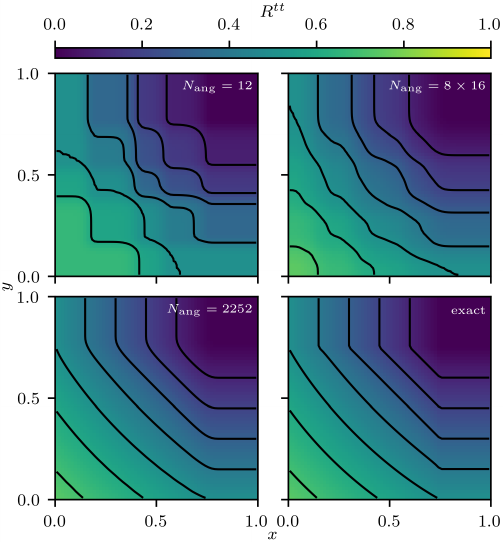}
  \caption{Solutions for the 2D hohlraum test with $64^2$ cells in space. The left panels show results using low-~and very-high-resolution geodesic grids, while the upper right panel shows a medium-resolution result on a latitude/longitude grid. Radiation is emitted isotropically from the left and bottom walls. Contours mark every $0.1$ units in energy density. We recover the correct behavior with minimal noise in the upper right panel using a reasonable number of angles. \label{fig:hohlraum_2d_solution}}
\end{figure}

We run this problem on a $64^2$ spatial grid. Again we use latitude/longitude grids with $N_\zetah$ varying from $2$ to $32$ and with $N_\psih = 2 N_\zetah$, as well as geodesic grids with $\Nang$ ranging from $12$ to $2252$. The results on three different grids are shown in Figure~\ref{fig:hohlraum_2d_solution}. The radiation passes through itself throughout the grid, avoiding artifacts characteristic of simple closure schemes \citep[see][for a comparison between methods]{Ryan2020}. This method also shows less noise compared to MOCMC using $\Nsamp = \Nang$ samples. The highest resolution shown is nearly indistinguishable from the exact solution, with only a slight smoothing of the sharp kinks seen in the exact contour levels, as is expected given the coarse spatial resolution.

\begin{figure}
  \centering
  \includegraphics{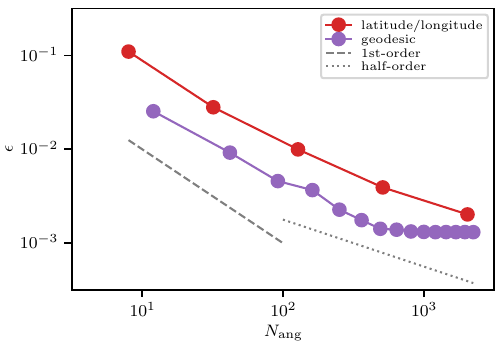}
  \caption{Convergence of the 2D hohlraum test with angular resolution. For all tests we use $64^2$ cells in space. As in the 1D case, convergence is asymptotically half order in the total number of angles, as expected. \label{fig:hohlraum_2d_convergence}}
\end{figure}

Figure~\ref{fig:hohlraum_2d_convergence} shows the convergence of our energy density to the exact solution with angular resolution. Here we define
\begin{equation}
  \epsilon = \int_0^1 \! \int_0^1 \abs[\big]{\Rcon{tt} - \Rcone{tt}} \, \dd x \, \dd y.
\end{equation}
As with the 1D hohlraum, we are essentially measuring the integral of a discontinuous function on the sphere using a finite number of angular bins. Thus we only expect half-order convergence, which we exceed at low resolutions and match at intermediate resolutions. Eventually, the error reaches the noise floor imposed by the spatial resolution.

\subsection{Radiating Disk}
\label{sec:tests:radiating_disk}

As an example of a hohlraum-like test in a curved spacetime, we consider the problem of a thin disk around a spinning black hole (mass $M$, $a = 15/16$), radiating into vacuum. Following \citet{Ryan2020}, we run a 2D simulation ($r$ and $\theta$) in which the midplane is treated as an internal boundary, with the radiation intensity there given by the Novikov--Thorne solution. That is, we calculate a physical temperature via
\begin{subequations} \begin{align}
  T & = (4 \times 10^7\ \K) \alpha^{-1/4} \parenpow{\bigg}{\frac{M}{3\ \Msun}}{-1/4} \notag \\
  & \quad \qquad \times r^{-3/8} A^{-1/2} B^{1/2} E^{1/4}, \\
  A & = 1 + \frac{a^2}{r^2} + \frac{2 a^2}{r^3}, \\
  B & = 1 + \parenpow{\bigg}{\frac{a^2}{r^3}}{1/2}, \\
  E & = 1 + \frac{4 a^2}{r^2} - \frac{4 a^2}{r^3} + \frac{3 a^4}{r^4},
\end{align} \end{subequations}
and then we demand
\begin{equation}
  \Ih = \frac{1}{4 \pi} \arad T^4
\end{equation}
in code units. Note that this means the radiation is isotropic in the tetrad frame (and thus the normal frame), not in some frame corotating with the disk.

The disk occupies the single midplane cell (out of $129$ in $\theta$), and is truncated at $r = 6$ and $r = 10$. We use spherical Kerr--Schild coordinates, with $128$ cells logarithmically spaced from just inside the horizon (a single cell) to $r = 20$. We choose $\alpha = 0.05$ and $M = 3\ \Msun$ to parameterize the disk, and initialize the simulation with vacuum everywhere at $t = 0$.

\begin{figure}
  \centering
  \includegraphics{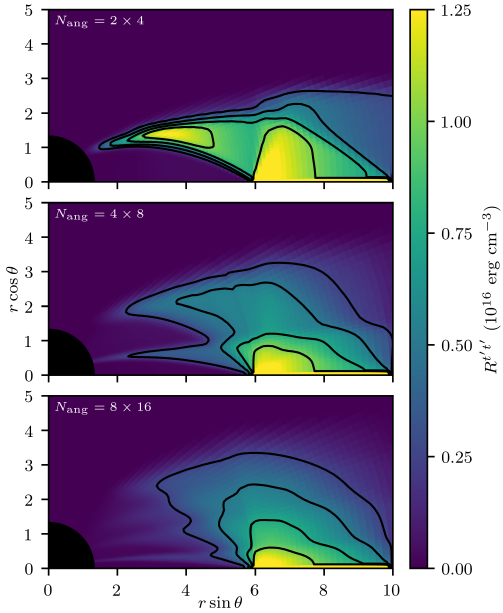}
  \caption{Normal-frame energy density at $t = 5$ in the radiating disk problem, with radiation sourced in the midplane in the region $6 \leq r \leq 10$. Contours are placed every $0.25$ units in energy density. Spatial resolution is fixed to $128 \times 129$, extending from inside the horizon to $r = 20$ and from pole to pole. With angular resolution increasing as indicated, the ray effects (narrow streams flowing into the black hole) diminish. \label{fig:radiating_disk}}
\end{figure}

Figure~\ref{fig:radiating_disk} shows the normal-frame (and tetrad-frame) radiation energy density at time $t = 5$, which is directly comparable to the results from MOCMC and other methods in \citet{Ryan2020}. The finite number of angles introduces some ray effects (the radiation streaming into the black hole should not be limited to discrete latitudes), which are exacerbated here by the choice of a tetrad aligned with the coordinates ($n^\twoh$ and $n^\threeh$ point in the $r$-~and $\theta$-directions everywhere). This effect is diminished as the angular resolution is increased, with the high-resolution result closely agreeing with the semi-analytic solution shown in \citet{Ryan2020}.

\subsection{Equilibration}
\label{sec:tests:equilibration}

Shifting attention from radiative transport to matter--radiation coupling, we perform an equilibration test to ensure that a fluid at rest will come to thermal equilibrium with an isotropic radiation field. Consider a static fluid with $\Gamma = 5/3$, $\rho = 1$, and initial value $\pgaszero = 2$, together with a coordinate-frame isotropic radiation field with initial energy density $\uradzero = 1$. We employ units in which $\kB / \mu \mprot$ and $\arad$ are unity, meaning the initial temperatures are $\Tgaszero = 2$ and $\Tradzero = 1$. We set scattering absorptivity $\alphabs$ to zero, but employ a nonzero $\alphaba = 0.1$.

\begin{figure}
  \centering
  \includegraphics{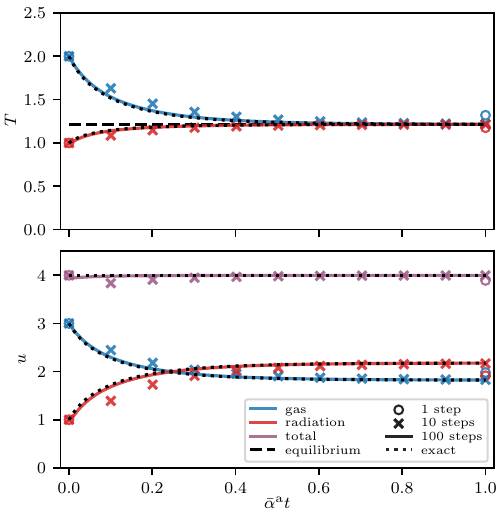}
  \caption{Evolution of temperature and energy in the static equilibration test taking timesteps of different sizes. The code values (colored) trace the exact solutions (dotted, black) as they evolve in time. The dashed line in the upper panel indicates the equilibrium temperature that should be approached. Time is measured in absorption coupling times. \label{fig:static_equilibration}}
\end{figure}

This scenario has no spatial dependence, though we run it on a Cartesian grid of $4^3$ cells to ensure this is reflected in the code. Larger cells allow larger hydrodynamical timesteps, and we adjust the grid in order to have $100$, $10$, or one timestep per coupling time $(\alphaba)^{-1}$. We use $5 \times 16$ angles with a Cartesian tetrad. The resulting temperatures and internal energies are plotted in Figure~\ref{fig:static_equilibration}. The temperatures approach the equilibrium value $\Tequil$ given by
\begin{equation}
  \frac{\Tequil \rho}{\Gamma - 1} + \Tequil^4 = \ugaszero + \uradzero,
\end{equation}
and the total energy is conserved. Moreover, the time evolution follows the exact result given by
\begin{align}
  \frac{\dd\ugas}{\dd t} & = \alphaba \paren[\Bigg]{\ugaszero + \uradzero \notag \\
  & \quad \qquad - \ugas - \parenpow{\bigg}{\frac{(\Gamma - 1) \ugas}{\rho}}{4}}.
\end{align}
The coupling is stable and approaches the correct equilibrium even with low time resolution, though the accuracy of the solution is reduced in such cases. We note that our implicit algorithm guarantees out-of-equilibrium temperatures approach one another but do not overshoot as a result of coupling.

We repeat this test, changing only the initial velocity of the fluid, now taken to be $u^1 = 1$. The initial radiation field is still isotropic in the coordinate frame. The matter and radiation should now come to thermal and momentum equilibrium. While the non-static version of the problem includes advection, there are no spatial gradients, and so we are still essentially testing coupling without transport. That is, we are reducing Equations~\eqref{eq:radiation_differential} and~\eqref{eq:grmhd_differential} to
\begin{subequations} \label{eq:moving_equilibration} \begin{align}
  & \frac{\dd}{\dd t} \paren[\big]{\rho \ucon{0}} = 0, \\
  & \frac{\dd}{\dd t} \paren[\big]{w \ucon{0} \ucon{0} - \pgas} = \notag \\
  & \qquad \alphaba \paren[\big]{\ucon{0} \Rcon{00} - \ucon{1} \Rcon{01} - T^4 \ucon{0}}, \\
  & \frac{\dd}{\dd t} \paren[\big]{w \ucon{0} \ucon{1}} = \notag \\
  & \qquad \alphaba \paren[\big]{\ucon{0} \Rcon{01} - \ucon{1} \Rcon{11} - T^4 \ucon{1}}, \\
  & \frac{\dd}{\dd t} I_n = \notag \\
  & \qquad \alphaba \paren[\bigg]{\frac{T^4}{4 \pi} \parenpow{\big}{\ucon{0} - \ucon{1} \naltconn{1}}{-3} - \paren[\big]{\ucon{0} - \ucon{1} \naltconn{1}} I_n},
\end{align} \end{subequations}
subject to the definitions
\begin{subequations} \begin{align}
  T & = \frac{\pgas}{\rho}, \\
  w & = \rho + \frac{\Gamma}{\Gamma - 1} \pgas, \\
  \ucon{0} & = \parenpow{\big}{1 + \ucon{1} \ucon{1}}{1/2}, \\
  \Rcon{\alpha\beta} & = \sum_n I_n \naltconn{\alpha} \naltconn{\beta} \Delta\Omega_n, \\
  \naltconn{0} & = 1.
\end{align} \end{subequations}

\begin{figure}
  \centering
  \includegraphics{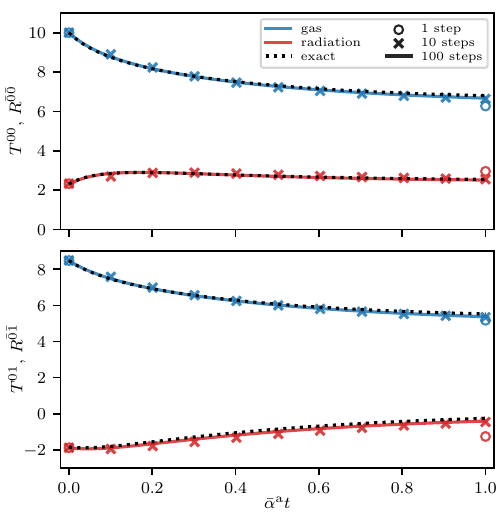}
  \caption{Evolution of energy and momentum densities in the moving equilibration test taking timesteps of different sizes. The code values (colored) trace the exact solutions (dotted, black) as they evolve in time. Time is measured in fluid-frame absorption coupling times. \label{fig:moving_equilibration}}
\end{figure}

Figure~\ref{fig:moving_equilibration} shows the coordinate-frame fluid energy and $x$-momentum densities, as well as the fluid-frame radiation energy and $x$-momentum densities, resulting from this test. The code values agree with the solution to the coupled ordinary differential equation of Equation~\eqref{eq:moving_equilibration}. Timesteps longer than approximately the coupling time suffer from an expected loss of accuracy but do not lead to catastrophic failures in the evolution. 

\subsection{Diffusion}
\label{sec:tests:diffusion}

In the limit of large scattering absorptivity $\alphaes \gg 1$ and vanishing absorption $\alphaea = 0$, radiation should diffuse through a background medium. If the medium is taken to be at rest and to affect but be unaffected by the radiation, the zeroth and first moments of the gray version of Equation~\eqref{eq:radiation_differential} reduce in this case to
\begin{subequations} \begin{align}
  \partial_t E - \frac{1}{3 \alphaes} \partial_x^2 E & \stackrel{\alphaes \to \infty}{\longrightarrow} 0, \\
  \partial_t \Fcon{x} & \stackrel{\phantom{\alphaes \to \infty}}{=} 0,
\end{align} \end{subequations}
where we assume $F^x = -(3 \alphaes)^{-1} \partial_x E$ together with the Eddington approximation for the second moments. Given an initial Gaussian profile with width $\sigma$, amplitude $\Epeak$, and offset $\Eoff$, the standard solution to the asymptotic equation is
\begin{subequations} \label{eq:diffusion} \begin{align}
  E & = \Epeak A \exp\paren[\bigg]{-\frac{A^2 x^2}{2 \sigma^2}} + \Eoff, \\
  A & = \parenpow{\bigg}{1 + \frac{2 t}{3 \alphaes \sigma^2}}{-1/2}.
\end{align} \end{subequations}

\begin{figure}
  \centering
  \includegraphics{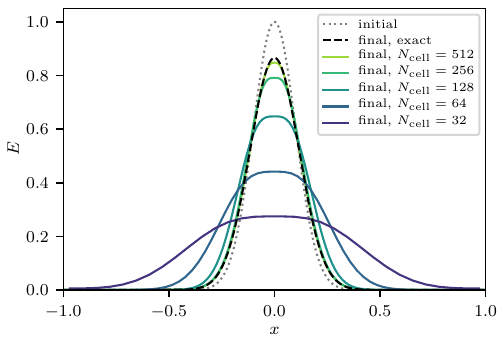}
  \caption{Initial and final radiation energy densities in the static diffusion problem. If spatial resolution is too low, numerical diffusion overwhelms physical diffusion set by scattering. The calculated solution approaches the exact one as resolution increases. Angular resolution is held constant with $1 \times 4$ bins. \label{fig:static_diffusion}}
\end{figure}

We initialize a 1D problem according to Equations~\eqref{eq:diffusion} at $t = 0$, fixing $\sigma = 0.1$, $\Epeak = 1$, $\Eoff = 0$, and $\alphaes = 10^4$. In order to initialize the intensities, we use the following prescription from \citet{Minerbo1978} in the fluid frame (which matches the coordinate frame here):
\begin{subequations} \label{eq:minerbo} \begin{align}
  \Ib & =
  \begin{dcases}
    \frac{1}{4 \pi} \paren[\big]{\Eb + 3 \nh_\ab F^\ab}, & f \leq \frac{1}{3}, \\
    \frac{1}{9 \pi} \cdot \frac{(2 - 3 f) f \Eb + \nh_\ab F^\ab}{f (1 - f)^2}, & f > \frac{1}{3};
  \end{dcases} \\
  f & = \frac{1}{\Eb} \parenpow{\big}{\Fcov{\ab} \Fcon{\ab}}{1/2}.
\end{align} \end{subequations}
The Eddington factor for this intensity field is exactly $1/3$ for $f \leq 1/3$. The fluid state is held constant throughout the calculation. Using $N_\zeta = 1$ and $N_\psi = 4$ angles and a variable number $\Ncell$ of spatial cells to cover $x \in [-1,\, 1]$, the radiation energy density at time $t = 50$ is shown in Figure~\ref{fig:static_diffusion}. As resolution increases, the pulse's diffusion approaches the expectation given $\alphaes$.

\begin{figure}
  \centering
  \includegraphics{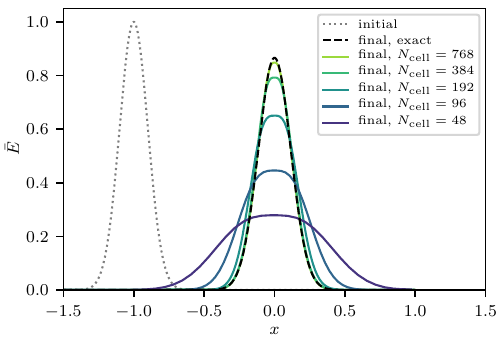}
  \caption{Initial and final fluid-frame radiation energy densities in the advected diffusion problem. The advection speed of the pulse is correctly captured even at low resolutions. High resolutions are needed in order to achieve the correct diffusion rate. Angular resolution is held constant with $1 \times 4$ bins. \label{fig:advected_diffusion}}
\end{figure}

We can repeat this problem in a frame in which the fluid is moving with velocity $v^x = 0.02$. The solution for all spacetime is given by Equations~\eqref{eq:diffusion}, with coordinate-frame $E$, $t$, and $x$ replaced by fluid-frame $\Eb$, $\tb$, and $\xb$, respectively. That is, the initial radiation field is no longer isotropic in the coordinate frame, and the initial conditions at $t = 0$ correspond to those at the surface $\tb = -v^x \xb$. The fluid-frame energy density at $t = 50$ is shown in Figure~\ref{fig:advected_diffusion}, where the initial pulse is shifted to the left by unity, and the simulation covers $x \in [-2,\, 1]$. The pulse always advects at the appropriate speed, and again increased resolution results in a closer match to the exact solution.

\begin{figure}
  \centering
  \includegraphics{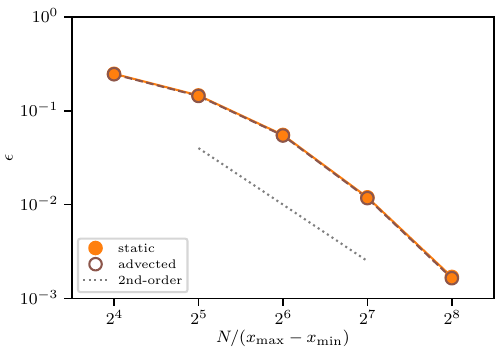}
  \caption{Convergence of the static and advected diffusion problems with spatial resolution. In all cases we use $1 \times 4$ angles. The solutions converge at approximately second order. \label{fig:diffusion}}
\end{figure}

We can quantify convergence via the $L^1$ norm of the difference between the calculated and exact fluid-frame radiation energy densities,
\begin{equation}
  \epsilon = \int_{\xmin}^{\xmax} \abs[\big]{\Eb_\mathrm{calc} - \Eb_\mathrm{exact}} \, \dd x.
\end{equation}
Figure~\ref{fig:diffusion} shows the errors as functions of the number of cells per unit length in $x$. The static and advected cases track each other closely, with convergence that is second order in spatial resolution for all but the lowest resolutions. This agrees with our use of PLM reconstruction and a second-order time integrator for this problem.

\subsection{Shocks}
\label{sec:tests:shocks}

A set of four standing, radiative shock solutions, first described by \citet{Farris2008}, have become standard tests for relativistic hydrodynamics codes. They consist of (1) a nonrelativistic shock, (2) a mildly relativistic shock, (3) a highly relativistic wave, and (4) a mildly relativistic wave, the last of which has radiation pressure dominating gas pressure. In principle, a code initialized with such an exact, stationary solution should hold that state, at least for several signal-crossing times. However, the solutions found by \citeauthor{Farris2008}\ assume the Eddington approximation in the fluid frame, which is not exactly true when full transport is considered.\footnote{See \citet{Tolstov2015} for a generalization to other closures, especially M1. Still, such solutions will not exactly match those obtained by full transport.}

Instead, we initialize the left and right halves of a 1D domain $[-20,\, 20]$ to the asymptotic states given by \citeauthor{Farris2008}\ (see their Table~1, where $\rho$, $\pgas$, and $\Eb$ implicitly define $\arad$ via thermal equilibrium, and where the fluid-frame flux is understood to vanish). The shock tubes are evolved, holding the boundary conditions at the asymptotic states, until they find the stationary solutions, which in all cases qualitatively match those given in the literature assuming the Eddington approximation. We choose the same run times ($40$, $400$, $80$, and $150$, respectively) and spatial resolution ($800$ cells) as in \citet{Ryan2020}. For the angular grid, we use $5 \times 10$ bins for the first two shocks, which is more than enough for those problems, and $30 \times 60$ bins for the latter two, whose large relative velocities require finely sampled angles.

\begin{figure}
  \centering
  \includegraphics{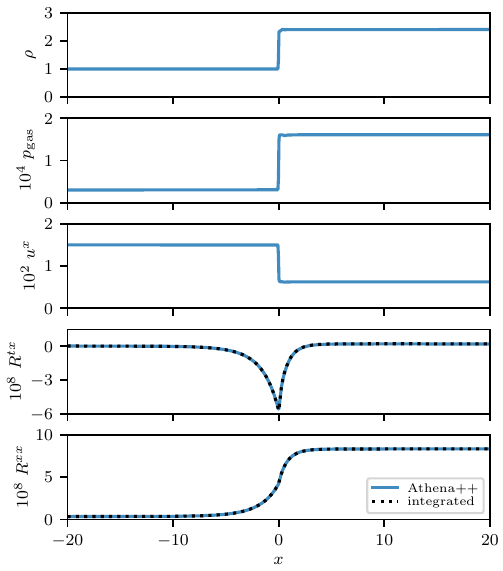}
  \caption{Nonrelativistic, gas-dominated, radiative shock, run with $800$ cells and $5 \times 10$ angles. \label{fig:shock_1}}
\end{figure}

\begin{figure}
  \centering
  \includegraphics{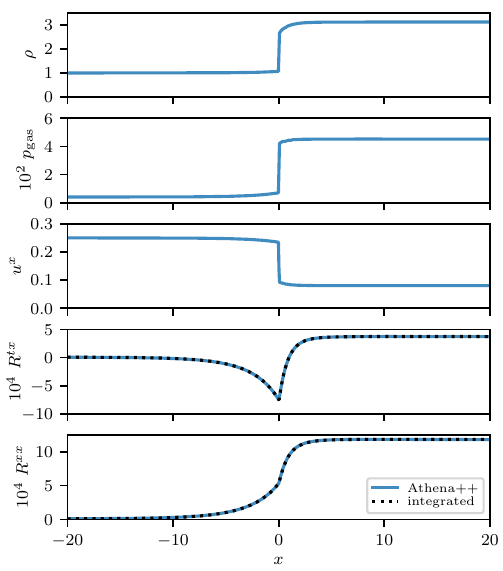}
  \caption{Mildly relativistic, gas-dominated, radiative shock, run with $800$ cells and $5 \times 10$ angles. \label{fig:shock_2}}
\end{figure}

\begin{figure}
  \centering
  \includegraphics{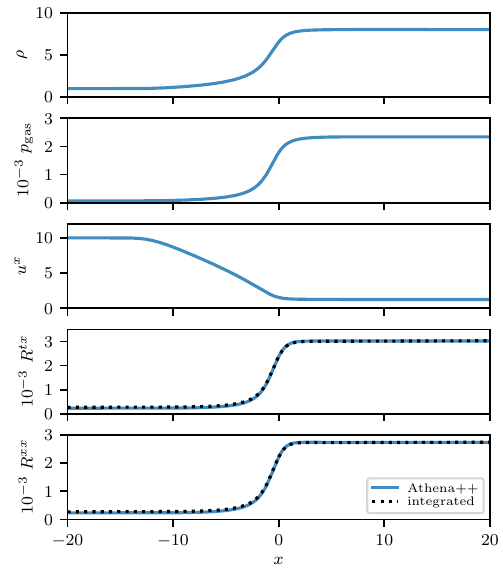}
  \caption{Highly relativistic, gas-dominated, radiative wave, run with $800$ cells and $30 \times 60$ angles. \label{fig:shock_3}}
\end{figure}

\begin{figure}
  \centering
  \includegraphics{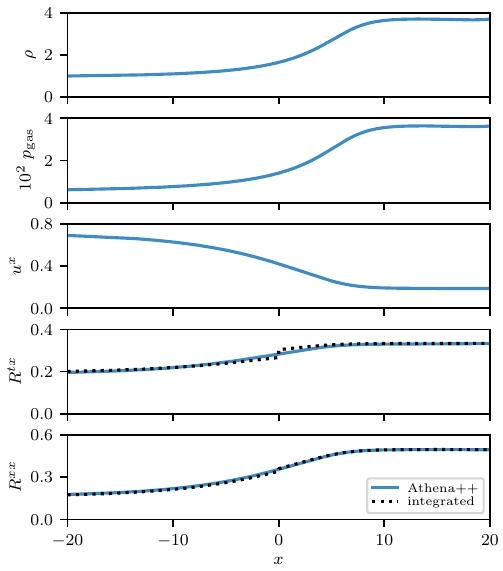}
  \caption{Mildly relativistic, radiation-dominated, radiative wave, run with $800$ cells and $5 \times 10$ angles. \label{fig:shock_4}}
\end{figure}

Following \citet{Ohsuga2016}, we further check the self-consistency of the numerical solutions by ensuring the radiation moments obey the appropriate time-independent equations. Given the hydrodynamical state ($\rho$, $\pgas$, and $u^x$) and fluid-frame moments $\Eb$ and $F^\xb$ found numerically across the domain, the steady-state, Minkowski, zeroth and first moments of Equation~\eqref{eq:radiation_differential} tell us
\begin{subequations} \begin{align}
  \frac{\dd}{\dd x} \Rcon{tx} & = -\Gcon{t} \notag \\
  & = -\kappaba \rho \paren[\big]{\Eb \ucon{t} - \arad T^4 \ucon{t} + \Fcon{\xb} \ucon{x}}, \\
  \frac{\dd}{\dd x} \Rcon{xx} & = -\Gcon{x} \notag \\
  & = -\kappaba \rho \paren[\big]{\Eb \ucon{x} - \arad T^4 \ucon{x} + \Fcon{\xb} \ucon{t}},
\end{align} \end{subequations}
where $T = \pgas / \rho$ and where $\kappaba$ is the absorption opacity (scattering is set to vanish). We can integrate the coordinate frame moments $R^{tx}$ and $R^{xx}$ from the boundaries to $x = 0$, comparing the resulting profiles with their numerical counterparts. Note the Eddington approximation does not enter into this process. Figures~\ref{fig:shock_1} through~\ref{fig:shock_4} show the results of the four shock tests. The integrated moments agree with the moments of the radiation field directly found by the code.

\subsection{Linear Waves}
\label{sec:tests:linear_waves}

As extremely stringent, quantitative tests of fluid dynamics codes, linear waves are essential to demonstrating that a code converges with resolution to the correct solution, and that convergence proceeds at the expected rate. Here we run a number of radiation-modified hydrodynamic and magnetohydrodynamic relativistic linear waves.

In all cases, we use a one-dimensional static background state with radiation in thermal and momentum equilibrium with the fluid. We perturb each of the $\Nprim$ primitive quantities $q_n$ defining this state, relative to their background values $q_{n,0}$, according to
\begin{equation} \label{eq:perturbation}
  q_n = q_{n,0} + \delta \cdot \Re \paren[\big]{e_n \exp(-\ii \omega t + \ii k x)},
\end{equation}
where $\delta = 10^{-4}$, $k = 2 \pi$, $t$ is initially $0$, and $\omega$ and $\{e_n\}$ are the complex eigenvalue and eigenvector corresponding to the desired wave (normalized such that $\sum_n \abs{e_n}^2 = 1$). At a later time $t$, the numerically computed state $\{Q_n\}$ is compared with the linear prediction given by Equation~\eqref{eq:perturbation}. The error is calculated according to
\begin{subequations} \begin{align}
  \epsilon & = \parenpow{\bigg}{\frac{1}{\Nprim'} \sum_n \epsilon_n^2}{1/2}, \\
  \epsilon_n & = \frac{1}{\delta} \int_0^1 \abs{Q_n - q_n} \, \dd x,
\end{align} \end{subequations}
where the sum is taken over all $\Nprim'$ quantities for which $e_n \neq 0$. We restrict our attention to waves with no vector components (background or perturbed) in one of the transverse directions ($z$).

The calculation of eigenvectors is detailed in Appendix~\ref{sec:linear_wave_analysis}, which provides solutions in terms of the MHD primitives and the fluid-frame radiation moments $\Eb$ and $F^\ab$. Intensities are computed from these moments according to Equations~\eqref{eq:minerbo}.

We use only four solid angles for these tests, divided by lines of longitude $\psih = 0,\, \pi / 2,\, \pi,\, 3 \pi / 2$. We fix the direction components to be $\nh^\zeroh = 1$, $\nh^\oneh = \pm 3^{-1/2}$, $\nh^\twoh = \pm 3^{-1/2}$, and $\nh^\threeh = 0$, meaning $\nh$ is not strictly null. In this way, however, a radiation field in 2D with equal intensity in all directions will be exactly Eddington:\ $\Rcon{\alpha\beta}$ will be proportional to $\diag(1,\, 1/3,\, 1/3,\, 0)$ rather than $\diag(1,\, 1/2,\, 1/2,\, 0)$. As an Eddington approximation is assumed in the derivation of the linear solutions, this is critical for ensuring the code is solving the same problem as the analytics.

Our grid, however, has $4$ intensity values in each cell, coupled via $3$ moments to the fluid (as there are no $z$-direction components). There is ambiguity in the intensity field in that one could add a given amount of $\hat{I}$ in the directions of $\psih = \pi / 4$ and $\psih = 5 \pi / 4$ while subtracting the same amount from the other two directions without changing $\Rcon{\zeroh\zeroh}$, $\Rcon{\zeroh\oneh}$, or $\Rcon{\zeroh\twoh}$. Similarly, for fixed fluid-frame moments $\Rcon{\zerob\zerob}$, $\Rcon{\zerob\oneb}$, and $\Rcon{\zerob\twob}$, there is a family of representable radiation fields with these moments, with the procedure of Section~\ref{sec:method:coupling} subtly choosing one solution. For the linear waves we enforce $\Rcon{\oneb\twob} = 0$ at the end of each matter--radiation coupling step, without modifying $\Rcon{\zerob\zerob}$, $\Rcon{\zerob\oneb}$, or $\Rcon{\zerob\twob}$, again ensuring we are making the same assumptions in the code as in the analytics.

For the hydrodynamic wave, we fix the fluid adiabatic index to be $\Gammagas = 5/3$, and we consider the three cases of a gas-dominated state ($\prad / \pgas = 1/10$), a radiation-dominated state ($\prad / \pgas = 10$), and a gas-radiation-equality state ($\prad / \pgas = 1$). All of our waves are run with constant absorption and scattering absorptivities $\alphaba = 10$ and $\alphabs = 10$. In the limit of infinite optical depth, the total adiabatic index of the gas and radiation in thermal equilibrium would correspond to the classical result
\begin{equation}
  \Gamma = \frac{32 (\prad / \pgas)^2 + 40 (\prad / \pgas) + 5}{24 (\prad / \pgas)^2 + 27 (\prad / \pgas) + 3},
\end{equation}
with associated sound speed
\begin{equation}
  \cs^2 = \frac{\Gamma (\pgas + \prad)}{\rho + \ugas + \urad + \pgas + \prad}.
\end{equation}
We choose to fix $\cs = 1/2$, which, together with an arbitrary normalization of $\rho = 1$, fixes the background state. The actual linear wavespeeds are slightly different from $1/2$ given our finite absorptivities. The exact background states and perturbations are cataloged in Appendix~\ref{sec:linear_wave_numerics}.

\begin{figure}
  \centering
  \includegraphics{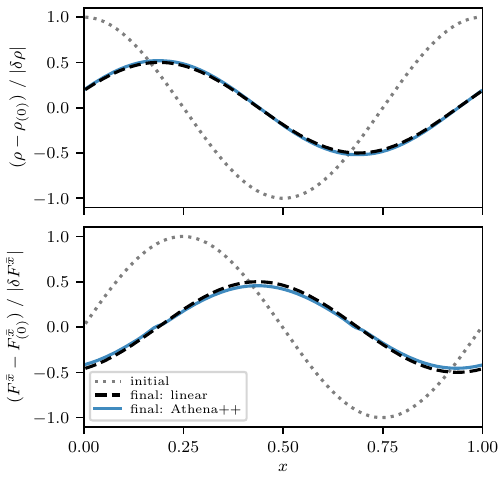}
  \caption{Linear wave solutions for the hydrodynamical sound wave with $\prad = \pgas$ at a resolution of $\Ncell = 128$. \label{fig:linear_wave_hydro_equal}}
\end{figure}

Each wave is run for a time of $-\log(2) / \Im(\omega)$, which will result in the initial linear perturbation amplitude decreasing by a factor of $2$. The initial and final perturbations in density and longitudinal fluid-frame radiation momentum, both according to linear theory and according to the code, are shown in Figure~\ref{fig:linear_wave_hydro_equal} for the case of $\prad = \pgas$ with $\Ncell = 128$ cells across the domain. The numerical solution closely matches linear theory for all variables.

\begin{figure}
  \centering
  \includegraphics{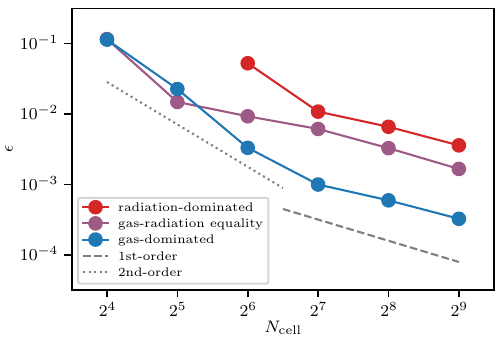}
  \caption{Convergence of the hydrodynamical sound wave for three different values of $\prad / \pgas$. We expect second-order convergence only when coupling is insignificant compared to advection; otherwise convergence should be first order. \label{fig:linear_wave_hydro}}
\end{figure}

The errors for the three hydrodynamic cases are shown in Figure~\ref{fig:linear_wave_hydro}. In all cases, we attain at least first-order convergence with sufficient resolution. Faster convergence reflects conditions where coupling errors (which are first order in time and second order in space) are dominated by advection errors (which are second order in time and space). At very low resolutions, the radiation-dominated case has order-unity errors, which is expected given the small number of angular bins we are using to represent the radiation field.

We also test a set of radiation-modified MHD waves. Here we consider the same three values of $\prad / \pgas$, keeping $\rho = 1$ and $\Gammagas = 5/3$ while adding the additional constraint $\pmag / \pgas = 1$. The optically thick limit of the longitudinal Alfv\'en speed is given by
\begin{subequations} \begin{align}
  \ca^2 & = \frac{B^x B^x}{\wgas + \wrad + \wmag}, \\
  \wgas & = \rho + \ugas + \pgas, \\
  \wrad & = \urad + \prad, \\
  \wmag & = B^x B^x + B^y B^y.
\end{align} \end{subequations}
By choosing a field inclination of $45^\circ$ ($B^x = B^y$, $B^x > 0$) and an Alv\'en speed of $\ca = 1/8$, we fix the background state, which conveniently separates all seven waves. The background states, as well as their linear eigenfunctions, are given in Appendix~\ref{sec:linear_wave_numerics}.

Figure~\ref{fig:linear_wave_mhd} summarizes the MHD results. We consider the rightgoing fast magnetosonic waves in all three backgrounds, achieving first-order convergence at high resolution as expected. We also show the results for a slow magnetosonic and an entropy wave in the $\prad = \pgas$ state, which again converge as expected. As magnetic fields do not directly couple to radiation, we forego the Alfv\'en waves.

\begin{figure}
  \centering
  \includegraphics{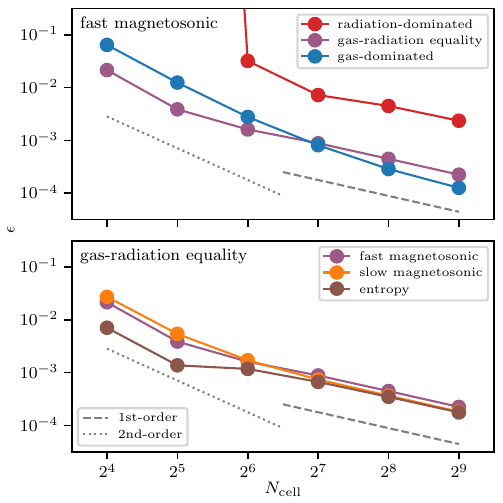}
  \caption{Convergence of the MHD waves. The top panel shows the fast magnetosonic wave for three different values of $\prad / \pgas$, and the bottom shows multiple waves for $\prad = \pgas$. The first-order convergence at high resolution is expected. \label{fig:linear_wave_mhd}}
\end{figure}

\subsection{Schwarzschild Atmosphere}
\label{sec:tests:schwarzschild_atmosphere}

In order to test radiation--fluid coupling in a nontrivial spacetime, we turn to the isothermal Schwarzschild atmosphere test presented in \citet{Ryan2015}. We consider an unmagnetized fluid around a nonspinning black hole with mass $M$ in Schwarzschild coordinates. Static, thermal equilibrium boundary conditions are imposed at inner and outer spherical shells, corresponding to a fixed temperature $\Tinf$ at infinity. This defines a spherically symmetric, thermal equilibrium, hydrostatic solution via $\nabla_\alpha (T^{\alpha\beta} + R^{\alpha\beta}) = 0$. Equivalently, the total pressure and internal energy must obey
\begin{equation} \label{eq:atmosphere}
  \frac{\dd\ptot}{\dd r} = -\frac{\rg (c^2 \rho + \utot + \ptot)}{r^2 \alpha^2},
\end{equation}
where $\alpha = (1 - 2 \rg / r)^{1/2}$ is the lapse and $\rg = G M / c^2$ is the gravitational radius.

For a fluid with constant adiabatic index $\Gamma$, the exact solution to Equation~\eqref{eq:atmosphere} can be obtained in a straightforward manner by considering the corresponding equation for $\dd (\alpha^{\Gamma / (\Gamma - 1)} \ptot) / \dd \alpha$, yielding
\begin{subequations} \begin{align}
  \ptot & = \frac{A}{\alpha^4} + B \paren[\bigg]{\ptotzero - \frac{A}{\alpha_0^4}}, \\
  A & = \frac{1}{3} \arad \Tinf^4, \\
  B & = \exp\paren[\bigg]{c^2 \frac{\mu \mprot}{\kB \Tinf} (\alpha_0 - \alpha)} \parenpow{\bigg}{\frac{\alpha_0}{\alpha}}{\Gamma / (\Gamma - 1)}.
\end{align} \end{subequations}
Here a $0$ subscript indicates the quantity is evaluated at the inner boundary. The remaining quantities can be found by noting the local temperature obeys
\begin{equation}
  T = \frac{\Tinf}{\alpha}.
\end{equation}

\begin{figure}
  \centering
  \includegraphics{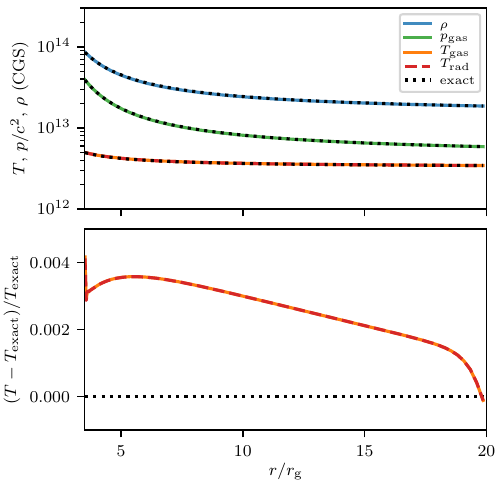}
  \caption{Solution for the 1D, spherically symmetric, isothermal, Schwarzschild atmosphere test. Here we use $128$ zones in radius and $64 \times 128$ angles for the radiation. \label{fig:atmosphere_solution}}
\end{figure}

Following \citet{Ryan2020}, we fix $M = \Msun$, a scale height parameter $H / r_0 = 1.6$ (which sets $\Tinf$), $\pradzero / \pgaszero$ = 43.5 (which sets $\rho_0$), and an optical depth parameter $\tau = 5$ (which sets the opacity, though the solution is independent of this value). A 1D simulation is initialized to the exact solution for $r \in [3.5,\, 20]$ with $128$ logarithmically spaced cells and run for $500$ gravitational times $\tg = \rg / c$. Figure~\ref{fig:atmosphere_solution} shows the resulting state, including the relative errors in gas and radiation temperatures, using $64 \times 128$ angles for the radiation.

\begin{figure}
  \centering
  \includegraphics{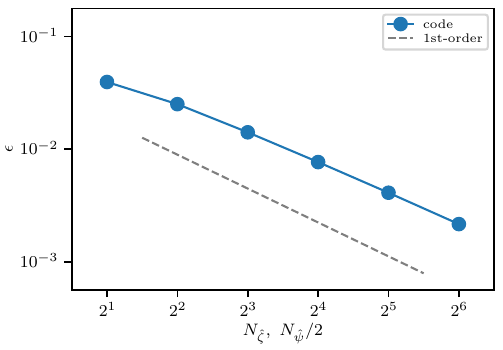}
  \caption{Convergence of the 1D, spherically symmetric, isothermal, Schwarzschild atmosphere test. Here we use $128$ zones in radius and vary the angular resolution. Errors are evaluated based on the gas and radiation temperatures deviating from the exact solution. \label{fig:atmosphere_convergence}}
\end{figure}

We further define the error in the solution to be
\begin{subequations} \begin{align}
  \epsilon & = \frac{1}{\sqrt{2}} \parenpow{\big}{\epsilon_\mathrm{gas}^2 + \epsilon_\mathrm{rad}^2}{1/2}, \\
  \epsilon_x & = \frac{1}{\rg^3 \Tinf} \int_{r_0}^{r_1} \abs{T_x - T_\mathrm{exact}} r^2 \, \dd r.
\end{align} \end{subequations}
Figure~\ref{fig:atmosphere_convergence} shows the run of this error with angular resolution, using grids with $2 \times 4$ to $64 \times 128$ angles. We find convergence that is first order in each of $N_\zetah$ and $N_\psih$.

\section{Black Hole Accretion Example}
\label{sec:example}

As a demonstration of the applicability of our algorithm to the physical systems we intend to study with it, we run a standard black hole accretion problem modified to include radiation. Here we use bars to denote quantities in code units, which are defined by a length scale $\rg$, a velocity scale $c$, a time scale $\tg = \rg / c$, a density scale $[\rho]$, and a temperature scale $\mu \mprot c^2 / \kB$ (so $\bar{T} = \bar{p} / \bar{\rho}$).

In the ubiquitous hydrodynamic equilibrium solution of \citet{Fishbone1976}, a torus around a spinning black hole is specified with two numbers, for example the radial coordinates of the inner edge of the torus, $\redgebar$, and the pressure maximum, $\rpeakbar$. From these, the fluid velocity and specific enthalpy $\bar{h} = 1 + \ugasbar / \bar{\rho} + \pgasbar / \bar{\rho}$ are determined everywhere in space. Given an arbitrary normalization $\bar{\rho} = \rhopeakbar$ at the pressure maximum, a gas with a fixed adiabatic index $\Gamma$, and an assumption of constant entropy ($\pgasbar / \bar{\rho}^\Gamma = \pgaspeakbar / \rhopeakbar^\Gamma$), the density and pressure are determined everywhere.

Here, we take the fluid to have the same velocity and specific enthalpy, but the latter is understood to be appropriate for an optically thick gas coupled to radiation:
\begin{align}
  \bar{h} & = 1 + \frac{\ugasbar}{\bar{\rho}} + \frac{\pgasbar}{\bar{\rho}} + \frac{\uradbar}{\bar{\rho}} + \frac{\pradbar}{\bar{\rho}} \notag \\
  & = 1 + \frac{\Gamma}{\Gamma - 1} \bar{T} + \frac{4 \aradbar}{3 \bar{\rho}} \bar{T}^4.
\end{align}
Setting $\rhopeakbar = 1$, this equation can be solved for the temperature $\Tpeakbar$ at the pressure maximum. For simplicity, we assume only the gamma-law gas has spatially constant entropy:
\begin{equation}
  \bar{T} \bar{\rho}^{1 - \Gamma} = \Tpeakbar \rhopeakbar^{1 - \Gamma}.
\end{equation}

For this demonstration, we initialize a small torus with $\redgebar = 20$ and $\rpeakbar = 41$ around a black hole with dimensionless spin $a = 15/16$. The density scale is chosen such that $\rhopeak = 5 \times 10^{-5}\ \g\ \cm^{-3}$ at the pressure maximum. The material's opacity is given by standard formulas for pure, ionized hydrogen:
\begin{subequations} \begin{align}
  \kappaa & = 8 \times 10^{22} \paren[\bigg]{\frac{\rho}{\g}} \parenpow{\bigg}{\frac{T}{\K}}{-3.5}\ \cm^2\ \g^{-1}, \\
  \kappas & = 0.4\ \cm^2\ \g^{-1}.
\end{align} \end{subequations}
The molecular weight is appropriately set to $\mu = 0.5$, and the adiabatic index is taken to be $\Gamma = 13/9$. The only remaining scale in the problem is $\rg$, set by taking the black hole to have a mass of $10\ \Msun$.

We add a standard weak poloidal magnetic field to the initial conditions, set from a vector potential $A_\phi \propto \max(\bar{\rho} - 0.2, 0)$. The normalization is chosen such that the ratio of the largest value of $\pmag$ to the largest value of $\pgas$ is $10^{-2}$. The result is that the average value of $\pmag / \pgas$, weighted by mass in regions with $\bar{\rho} > 0.2$, is $4.64 \times 10^{-2}$.

Here we use a Cartesian Kerr--Schild $(x,\, y,\, z)$ grid with $256^3$ cells at root level covering the interval $[-1024,\, +1024]$ in each direction. We add seven levels of static mesh refinement. The first doubles the resolution in all dimensions inside $[-512,\, +512]^3$; the next does so again inside $[-256,\, +256]^3$; and so on. The highest level attains a resolution of $16$ cells per $\rg$ throughout $[-8,\, +8]^3$. The radiation field is discretized with an $\Nang = 92$ geodesic grid.

\begin{figure*}
  \centering
  \includegraphics{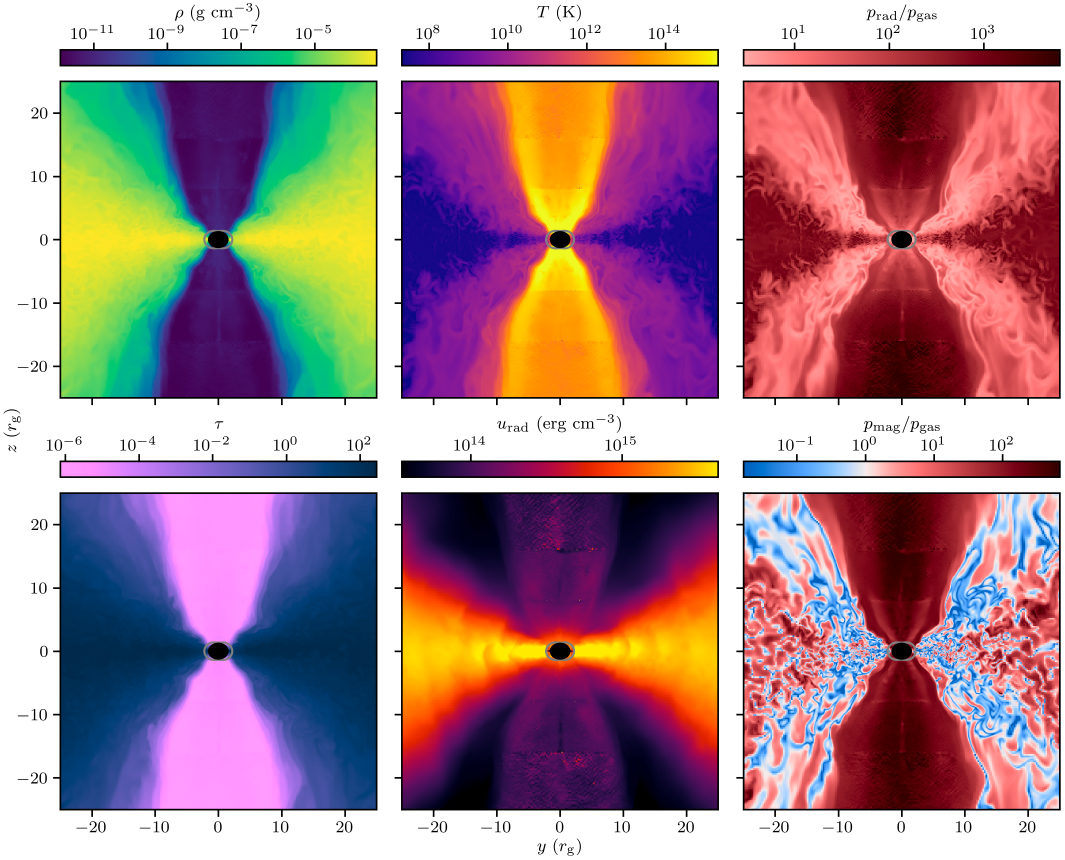}
  \caption{Slices ($x = 0$) through a 3D simulation of black hole accretion initialized from a standard magnetized torus augmented with radiation. Snapshots are taken at time $t = 10^4\ \tg$. The simulation domain extends inside the horizon, though here the black hole interior is masked. The gray line denotes the ergosphere. \label{fig:torus}}
\end{figure*}

Figure~\ref{fig:torus} shows slices through the simulation at time $10^4\ \tg$. Much of the disk has cooled below $10^8\ \K$. Radiation pressure dominates over gas pressure by factors of $10$ to $10^4$ everywhere, while magnetic pressure is comparable to gas pressure, with large fluctuations, throughout most of the domain. In one panel we show the optical depth per gravitational length,
\begin{equation}
  \tau = (\kappaa + \kappas) \rho \rg,
\end{equation}
illustrating that the bulk of the disk is optically thick, while the jet is optically thin.

\section{Performance}
\label{sec:performance}

For small numbers of angles (less than approximately eight), the per-cell cost of our GR radiation scheme is comparable to the per-cell cost of GRMHD without radiation, both in terms of computation and memory usage. The cost of radiation scales linearly with the number of angles; the only potential quadratic dependence comes from implicit coupling, though this is elided via the method discussed in Section~\ref{sec:method:coupling}.

\begin{figure}
  \centering
  \includegraphics{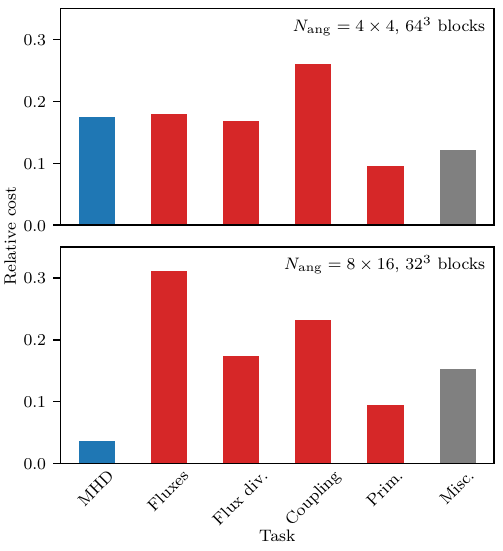}
  \caption{Relative costs associated with radiation using \code{Athena++} on a CPU node (Intel Ice Lake). We partition the costs into key tasks from Figure~\ref{fig:tasks}, with all fluid and magnetic field tasks bundled into ``MHD'' and all non-enumerated tasks treated as miscellaneous. \label{fig:timings_cpu}}
\end{figure}

Quantitatively, we measure the cost on a modern CPU node consisting of $64$ Ice Lake (Intel Xeon Platinum 8362) cores. Here we run a 3D linear wave with both magnetic fields and radiation using \code{Athena++}. In one test, we use a latitude/longitude grid with $4 \times 4$ angles, and with a domain decomposed into $64$ blocks of $64^3$ cells each. In another test, we employ $8 \times 16$ angles and $64$ blocks of $32^3$ cells each. These tests are distributed across an entire node via MPI, and the contributions of key tasks are measured. Figure~\ref{fig:timings_cpu} shows the relative costs, categorized as non-radiative MHD, calculation of radiation fluxes, updating radiation with flux divergences (in both space and angle), coupling radiation to matter, finding radiation primitives from conserved quantities, and all other tasks (e.g., communication).

Scientific production runs investigating black hole accretion will likely use an intermediate number of angles between the $16$ and $128$ shown in Figure~\ref{fig:timings_cpu}. Relative to GRMHD simulations, radiation--GRMHD can be expensive, but not overwhelmingly so, with our algorithm. For these two tests in particular, \code{Athena++} achieves a performance of $7.4 \times 10^6$ and $1.4 \times 10^6$ cell updates per second per node.

However, the accretion flows that require radiation to be simulated are generally those in which the fluid can cool enough to form a thin disk. Capturing three-dimensional turbulent processes in such systems may require extreme resolutions and thus exascale computers, which generally employ GPUs. We therefore implement our algorithm in the newly developed \code{AthenaK} code (Stone et al., in prep.), which employs Kokkos \citep{Carter2014} in order to run on both CPUs and GPUs.

\begin{figure}
  \centering
  \includegraphics{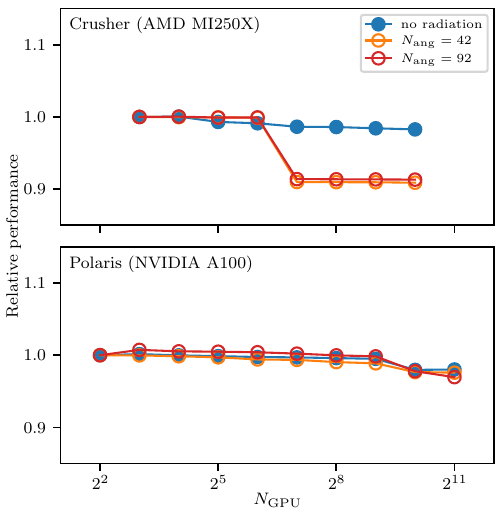}
  \caption{Weak scaling of radiation using \code{AthenaK} on GPU machines Crusher (with eight GPUs on four accelerators on each node) and Polaris (with four GPUs per node). Data is shown for GR radiation--hydrodynamics using either a level~2 or level~3 geodesic grid, and for GRMHD without radiation for comparison. Measured values are cell updates per second per GPU, normalized to the single-node values on the left end of each curve. We see nearly perfect scaling on Polaris out to $2048$ GPUs. \label{fig:scaling_gpu}}
\end{figure}

We measure scaling of the algorithm as implemented in \code{AthenaK} on two GPU machines. Crusher at Oak Ridge National Laboratory consists of AMD~MI250X accelerators, with each node containing four accelerators and each accelerator containing two GPUs. Polaris at Argonne National Laboratory consists of four NVIDIA~A100 GPUs per node. In each case we run a problem in Minkowski coordinates (exercising the full machinery of GR). Figure~\ref{fig:scaling_gpu} shows the scaling of a GRMHD problem with no radiation, together with that of a GR radiation--hydrodynamics problem using either $42$ or $92$ angles (geodesic grid level $2$ or $3$), reflecting typical angular resolutions we expect to use in practice. In these tests, the domain is decomposed into blocks of $128^3$, $128^3$, and $64^3$ cells, respectively.

On a single node on Crusher, we achieve absolute performances of $54.4$, $13.8$, and $5.89$ million cell updates per second per GPU for the three tests. On Polaris, we find $113$, $27.6$, and $12.6$ million cell updates per second per GPU.

The code shows excellent weak scaling out as far as tested with over $2000$ GPUs. Relative to a single node, both radiation cases show per-node performance of greater than $90\%$ on $128$ Crusher nodes, with non-radiative GRMHD over $98\%$ efficient, indicating further performance tuning on this particular machine can likely improve scaling. On Polaris, per-node performance on $512$ nodes is at least $97\%$ that of one node in all cases. With $42$ angles, the code runs at approximately $25\%$ the speed of a non-radiative calculation; this factor is $11\%$ with $92$ angles.

The radiation--GRMHD example given in Section~\ref{sec:example} was run to a time of $10^4\ \tg$ on $152$ A100 GPUs on Polaris in $26.7$ hours, corresponding to $4.37$ million cell updates (including $402$ million radiation angle updates) per second per GPU. Performance can be improved at the expense of wall time by using fewer GPUs, as each GPU here only evolved three blocks of $64^3$ cells, and as such was not always kept maximally busy performing computations.

\section{Summary}
\label{sec:summary}

We have implemented a finite-solid-angle method for radiation in the \code{Athena++} framework, as well as the newly developed, GPU-capable \code{AthenaK} code. This allows radiation to be added to finite-volume GRMHD simulations, with the intensity field in each cell discretized in angle. The radiation, fluid, and electromagnetic fields are all evolved on arbitrary stationary spacetimes, and they couple to one another as is physically appropriate. Section~\ref{sec:method} details the numerical algorithms we develop for this evolution and coupling. Along the way, this formalism allows flat-spacetime simulations in which non-Cartesian spatial dimensions have been suppressed (e.g., 2D axisymmetric simulations).

The complexity of the differential equations being solved, Equations~\eqref{eq:radiation_differential} and~\eqref{eq:grmhd_differential}, demands care be taken to write a robust and accurate code. We perform a large suite of tests (Section~\ref{sec:tests}), exercising all terms in the equations and all components of the method in multiple ways, to ensure that our implementation is calculating the correct solutions.

We look forward to using this code on a variety of astrophysical problems, especially black hole accretion near the Eddington rate. As shown in Section~\ref{sec:example}, the code is ready to simulate such systems. In contrast to earlier studies using M1 methods, our radiation fields are not restricted to be axisymmetric and monomodal in angle at every point. This can be important, especially if radiation pressure dominates over gas pressure in the optically thin--thick transition region near the disk surface. Unlike Monte Carlo methods, our procedure is naturally efficient at all optical depths, with a computational cost that remains fixed as the system evolves.

Radiation--GRMHD is never computationally cheap, and our method is no exception, especially if we choose to use many angles. Still, we have demonstrated the feasibility of running such problems on large machines, where our code has nearly perfect weak scaling (Section~\ref{sec:performance}). Crusher and Polaris are forerunners to Frontier and Aurora in the next generation of exascale machines, and we anticipate being able to address a large number of previously intractable questions with resources such as these. For example, we can now begin to study the ways in which radiation affects the creation of coronae, the launching of jets, the transport of mass and angular momentum in different magnetic regimes, and the torques in misaligned systems.

Our method naturally allows for discretization of radiation in energy (frequency) space separate from other dimensions. In fact, we only need to restore the gravitational redshift term present in Equation~\eqref{eq:radiation_integral_frequency} but not Equation~\eqref{eq:radiation_integral} to add frequency dependence. The only complication is that in many astrophysical systems, scattering processes can be nonlocal in energy, necessitating complicated implicit coupling in radiation momentum space, as for example treated in flat spacetime in \citet{Jiang2022}. Computational costs will necessarily scale with the number of energy bins. The extension of our general-relativistic, finite-solid-angle method to frequency dependence will be explored in a future work, together with appropriate tests of that functionality.

\acknowledgments

We are grateful to K.~Felker and P.~Grete for providing scaling data, and to O.~Blaes, J.~C.~Dolence, J.~M.~Miller, B.~R.~Ryan, and G.~N.~Wong for useful discussions and suggestions.

The authors acknowledge support from NASA TCAN grant 80NSSC21K0496, and SWD acknowledges support from NASA ATP grant 80NSSC18K1018.

This work made use of the following computing resources:\ the Extreme Science and Engineering Discovery Environment (XSEDE) cluster \code{Stampede2} at the Texas Advanced Computing Center (TACC), through allocation AST200005; the Oak Ridge Leadership Computing Facility cluster \code{Crusher} at the Oak Ridge National Laboratory, supported by the Office of Science of the U.~S. Department of Energy; the Argonne Leadership Computing Facility cluster \code{Polaris} at Argonne National Laboratory, supported by the Office of Science of the U.~S. Department of Energy; the clusters \code{Rusty} and \code{Popeye}, supported by the Scientific Computing Core at the Flatiron Institute, a division of the Simons Foundation; and the Princeton Research Computing cluster \code{Stellar}, managed and supported by the Princeton Institute for Computational Science and Engineering (PICSciE) and the Office of Information Technology's High Performance Computing Center and Visualization Laboratory at Princeton University.

This work has been assigned a document release number LA-UR-23-21167.

\bibliographystyle{aasjournal}
\bibliography{references}

\appendix

\section{Linear Wave Analysis}
\label{sec:linear_wave_analysis}

Consider a fluid with constant adiabatic index $\Gamma = \Gammagas$, longitudinal magnetic field $B^1$, energy-mean absorption coefficient $\alphabea$, Planck-mean absorption coefficient $\alphabpa$, and flux-mean absorption-plus-scattering coefficient $\alphabf$. Let the set of primitives
\begin{equation}
  \prim = \braces[\big]{\rho, \pgas, \ucon{1}, \ucon{2}, \ucon{3}, \Bcon{2}, \Bcon{3}, \Eb, \Fcon{\oneb}, \Fcon{\twob}, \Fcon{\threeb}}
\end{equation}
be indexed by $m,\, n \in \{1,\, \ldots,\, 11\}$. For convenience, define the derived quantities
\begin{subequations} \begin{align}
  \ucon{0} & = \parenpow{\big}{1 + \ucov{a} \ucon{a}}{1/2}, \\
  \bcon{0} & = \ucov{a} \Bcon{a}, \\
  \bcon{a} & = \frac{1}{u^0} \paren[\big]{\Bcon{a} + \bcon{0} \ucon{a}}, \\
  \pmag & = \frac{1}{2} \paren[\big]{-\bcon{0} \bcon{0} + \bcov{a} \bcon{a}}, \\
  w & = \rho + \frac{\Gamma}{\Gamma - 1} \pgas + 2 \pmag, \\
  T & = \frac{\pgas}{\rho},
\end{align} \end{subequations}
where spatial indices can be raised or lowered at will, and the last equation implies a factor of $\kB / \mu \mprot$ absorbed into the radiation constant $\arad$.

The 1D equations of special-relativistic radiation MHD can be written as
\begin{subequations} \label{eq:sr_1d} \begin{align}
  \pp{0} \paren[\big]{\rho \ucon{0}} + \pp{1} \paren[\big]{\rho \ucon{1}} & = 0, \\
  \pp{0} \paren[\big]{w \ucon{0} \ucon{\alpha} - \pgas \kroncon{0\alpha} - \pmag \kroncon{0\alpha} - \bcon{0} \bcon{\alpha}} + \pp{1} \paren[\big]{w \ucon{1} \ucon{\alpha} + \pgas \kroncon{1\alpha} + \pmag \kroncon{1\alpha} - \bcon{1} \bcon{\alpha}} & = \lorentz{\alpha}{\betab} \Gcon{\betab}, \\
  \pp{0} \Bcon{2} + \pp{1} \paren[\big]{\bcon{[2} \ucon{1]}} & = 0, \\
  \pp{0} \Bcon{3} + \pp{1} \paren[\big]{\bcon{[3} \ucon{1]}} & = 0, \\
  \pp{0} \paren[\big]{\lorentz{0}{\gammab} \lorentz{\alphab}{\deltab} \Rcon{\gammab\deltab}} + \pp{1} \paren[\big]{\lorentz{1}{\gammab} \lorentz{\alpha}{\deltab} \Rcon{\gammab\deltab}} & = -\lorentz{\alphab}{\betab} \Gcon{\betab}.
\end{align} \end{subequations}
Here the Lorentz boost is
\begin{subequations} \begin{align}
  \lorentz{0}{\zerob} & = \ucon{0}, \\
  \lorentz{a}{\zerob} & = \ucon{a}, \\
  \lorentz{0}{\bb} & = \ucov{b}, \\
  \lorentz{a}{\bb} & = \frac{u^a u_b}{1 + u^0} + \kronsplit{a}{b},
\end{align} \end{subequations}
and the coupling force is
\begin{subequations} \begin{align}
  \Gcon{\zerob} & = \alphabea \Eb - \alphabpa \arad T^4, \\
  \Gcon{\ab} & = \alphabf \Fcon{\ab}.
\end{align} \end{subequations}
We assume the Eddington closure in the fluid frame, so the radiation stress--energy tensor has components
\begin{subequations} \begin{align}
  \Rcon{\zerob\zerob} & = \Eb, \\
  \Rcon{\zerob\ab},\, \Rcon{\ab\zerob} & = \Fcon{\ab}, \\
  \Rcon{\ab\bb} & = \frac{1}{3} \Eb \kroncon{ab}.
\end{align} \end{subequations}

Let $\prim$ specify a stationary and homogeneous state. In order for Equations~\eqref{eq:sr_1d} to hold, it is then necessary and sufficient that $G^\alphab$ vanish. If none of the absorptivities vanish, this is equivalent to requiring
\begin{subequations} \begin{align}
  \Eb & = \frac{\alphabpa}{\alphabea} \arad T^4, \\
  \Fcon{\ab} & = 0,
\end{align} \end{subequations}
and we will assume these relations always hold.

Equations~\eqref{eq:sr_1d} are in flux--conservative form
\begin{equation}
  \pp{0} \cons^m + \pp{1} \flux^m = \source^m.
\end{equation}
Linearizing with respect to perturbed primitives, with perturbations of the form
\begin{equation}
  \delta\prim^n = \Re \paren[\Big]{\delta\prim_0^n \exp\paren[\big]{-\ii \omega t + \ii k x}}
\end{equation}
for unknown constants $\delta\prim_0^n$, we arrive at the generalized eigenvalue problem
\begin{equation} \label{eq:eigenvalue}
  \paren[\bigg]{k \pfrac{\flux^m}{\prim^n} + \ii \pfrac{\source^m}{\prim^n}} \delta\prim_0^n = \omega \pfrac{\cons^m}{\prim^n} \delta\prim_0^n
\end{equation}
for complex eigenvalues $\omega$ and eigenvectors $\{\delta\prim_0^n\}$.

The matrices of derivatives in Equation~\eqref{eq:eigenvalue} are given by
\begin{subequations} \begin{align}
  \pfrac{\cons^m}{\prim^n} & =
  \begin{pmatrix}
    \ucon{0} & 0 & \frac{1}{u^0} \rho \ucov{b} & 0 & 0 & 0 & \zerocov{b} \\
    \ucon{0} \ucon{0} & \pfrac{\cons^2}{\prim^2} & \pfrac{\cons^2}{\prim^{b+2}} & \pfrac{\cons^2}{\prim^6} & \pfrac{\cons^2}{\prim^7} & 0 & \zerocov{b} \\
    \ucon{0} \ucon{a} & \frac{\Gamma}{\Gamma - 1} \ucon{0} \ucon{a} & \pfrac{\cons^{a+2}}{\prim^{b+2}} & \pfrac{\cons^{a+2}}{\prim^6} & \pfrac{\cons^{a+2}}{\prim^7} & \zerocon{a} & \zerosplit{a}{b} \\
    0 & 0 & \zerocov{b} & 1 & 0 & 0 & \zerocov{b} \\
    0 & 0 & \zerocov{b} & 0 & 1 & 0 & \zerocov{b} \\
    0 & 0 & \frac{8}{3} \Eb \ucov{b} & 0 & 0 & \pfrac{\cons^8}{\prim^8} & 2 \ucon{0} \ucov{b} \\
    \zerocon{a} & \zerocon{a} & \pfrac{\cons^{a+8}}{\prim^{b+2}} & \zerocon{a} & \zerocon{a} & \frac{4}{3} \ucon{0} \ucon{a} & \pfrac{\cons^{a+8}}{\prim^{b+8}}
  \end{pmatrix}, \\
  \pfrac{\cons^2}{\prim^2} & = \frac{\Gamma}{\Gamma - 1} \ucon{0} \ucon{0} - 1, \\
  \pfrac{\cons^2}{\prim^{b+2}} & = 2 \paren[\bigg]{\rho + \frac{\Gamma}{\Gamma - 1} \pgas} \ucov{b} - \frac{1}{u^0 u^0} \paren[\Big]{\bcon{0} \Bcov{b} - 2 \pmag \ucov{b}}, \\
  \pfrac{\cons^2}{\prim^6} & = 2 \Bcon{2} - \frac{1}{u^0} \bcon{2}, \\
  \pfrac{\cons^2}{\prim^7} & = 2 \Bcon{3} - \frac{1}{u^0} \bcon{3}, \\
  \pfrac{\cons^{a+2}}{\prim^{b+2}} & = \frac{1}{u^0} \paren[\bigg]{\paren[\Big]{w \ucon{0} \ucon{0} - \bcon{0} \bcon{0}} \kronsplit{a}{b} + (w - 4 \pmag) \ucon{a} \ucov{b} + \bcon{a} \bcov{b} - \Bcon{a} \Bcov{b} - \frac{1}{u^0} \bcon{a} \Bcov{b}}, \\
  \pfrac{\cons^{a+2}}{\prim^6} & = \ucon{[a} \bcon{2]} + \frac{1}{u^0} \paren[\Big]{\ucon{a} \Bcon{2} - \bcon{0} \kronsplit{a}{2}}, \\
  \pfrac{\cons^{a+2}}{\prim^7} & = \ucon{[a} \bcon{3]} + \frac{1}{u^0} \paren[\Big]{\ucon{a} \Bcon{3} - \bcon{0} \kronsplit{a}{3}}, \\
  \pfrac{\cons^8}{\prim^8} & = \frac{4}{3} \ucon{0} \ucon{0} - \frac{1}{3}, \\
  \pfrac{\cons^{a+8}}{\prim^{b+2}} & = \frac{4}{3} \Eb \paren[\bigg]{\frac{1}{u^0} \ucon{a} \ucov{b} + \ucon{0} \kronsplit{a}{b}}, \\
  \pfrac{\cons^{a+8}}{\prim^{b+8}} & = \frac{1 + 2 u^0}{1 + u^0} \ucon{a} \ucov{b} + \ucon{0} \kronsplit{a}{b};
\end{align} \end{subequations}
\begin{subequations} \begin{align}
  \pfrac{\flux^m}{\prim^n} & =
  \begin{pmatrix}
    \ucon{1} & 0 & \rho \kronsplit{1}{b} & 0 & 0 & 0 & \zerocov{b} \\
    \ucon{0} \ucon{1} & \frac{\Gamma}{\Gamma - 1} \ucon{0} \ucon{1} & \pfrac{\flux^2}{\prim^{b+2}} & \pfrac{\flux^2}{\prim^6} & \pfrac{\flux^2}{\prim^7} & 0 & \zerocov{b} \\
    \ucon{1} \ucon{a} & \pfrac{\flux^{a+2}}{\prim^2} & \pfrac{\flux^{a+2}}{\prim^{b+2}} & \pfrac{\flux^{a+2}}{\prim^6} & \pfrac{\flux^{a+2}}{\prim^7} & \zerocon{a} & \zerosplit{a}{b} \\
    0 & 0 & \pfrac{\flux^6}{\prim^{b+2}} & \frac{1}{u^0} \ucon{1} & 0 & 0 & \zerocov{b} \\
    0 & 0 & \pfrac{\flux^7}{\prim^{b+2}} & 0 & \frac{1}{u^0} \ucon{1} & 0 & \zerocov{b} \\
    0 & 0 & \pfrac{\flux^8}{\prim^{b+2}} & 0 & 0 & \frac{4}{3} \ucon{0} \ucon{1} & \pfrac{\flux^8}{\prim^{b+8}} \\
    \zerocon{a} & \zerocon{a} & \frac{4}{3} \Eb \ucon{(1} \kronsplit{a)}{b} & \zerocon{a} & \zerocon{a} & \pfrac{\flux^{a+8}}{\prim^8} & \pfrac{\flux^{a+8}}{\prim^{b+8}}
  \end{pmatrix}, \\
  \pfrac{\flux^2}{\prim^{b+2}} & = \frac{1}{u^0} \paren[\bigg]{\paren[\Big]{w \ucon{0} \ucon{0} - \bcon{0} \bcon{0}} \kronsplit{1}{b} + (w - 4 \pmag) \ucon{1} \ucov{b} + \bcon{1} \bcov{b} - \Bcon{1} \Bcov{b} - \frac{1}{u^0} \bcon{1} \Bcov{b}}, \\
  \pfrac{\flux^2}{\prim^6} & = \ucon{[1} \bcon{2]} + \frac{1}{u^0} \ucon{1} \Bcon{2}, \\
  \pfrac{\flux^2}{\prim^7} & = \ucon{[1} \bcon{3]} + \frac{1}{u^0} \ucon{1} \Bcon{3}, \\
  \pfrac{\flux^{a+2}}{\prim^2} & = \frac{\Gamma}{\Gamma - 1} \ucon{1} \ucon{a} + \kroncon{1a}, \\
  \pfrac{\flux^{a+2}}{\prim^{b+2}} & = \frac{1}{u^0 u^0} \paren[\bigg]{w \ucon{0} \ucon{0} \ucon{(1} \kronsplit{a)}{b} + \paren[\Big]{2 \ucon{1} \ucon{a} + \kroncon{1a}} \paren[\Big]{\bcon{0} \Bcov{b} - 2 \pmag \ucov{b}} + 2 \ucov{b} \bcon{1} \bcon{a} - \ucon{0} \bcon{0} \bcon{(1} \kronsplit{a)}{b} - \ucon{0} \ucon{(1} \bcon{a)} \Bcov{b}}, \\
  \pfrac{\flux^{a+2}}{\prim^6} & = \frac{1}{u^0} \paren[\bigg]{\bcon{2} \paren[\Big]{2 \ucon{1} \ucon{a} + \kroncon{1a}} - \ucon{2} \ucon{(1} \bcon{a)} - \bcon{1} \kroncon{2a}}, \\
  \pfrac{\flux^{a+2}}{\prim^7} & = \frac{1}{u^0} \paren[\bigg]{\bcon{3} \paren[\Big]{2 \ucon{1} \ucon{a} + \kroncon{1a}} - \ucon{3} \ucon{(1} \bcon{a)} - \bcon{1} \kroncon{3a}}, \\
  \pfrac{\flux^6}{\prim^{b+2}} & = \frac{1}{u^0} \Bcon{[2} \kronsplit{1]}{b} + \frac{1}{u^0 u^0} \bcon{[1} \ucon{2]}, \\
  \pfrac{\flux^7}{\prim^{b+2}} & = \frac{1}{u^0} \Bcon{[3} \kronsplit{1]}{b} + \frac{1}{u^0 u^0} \bcon{[3} \ucon{1]}, \\
  \pfrac{\flux^8}{\prim^{b+2}} & = \frac{4}{3} \Eb \paren[\bigg]{\frac{1}{u^0} \ucon{1} \ucov{b} + \ucon{0} \kronsplit{1}{b}}, \\
  \pfrac{\flux^8}{\prim^{b+8}} & = \frac{1 + 2 u^0}{1 + u^0} \ucon{1} \ucov{b} + \ucon{0} \kronsplit{1}{b}, \\
  \pfrac{\flux^{a+8}}{\prim^8} & = \frac{1}{3} \kroncon{1a} + \frac{4}{3} \ucon{1} \ucon{a}, \\
  \pfrac{\flux^{a+8}}{\prim^{b+8}} & = \ucon{(1} \kronsplit{a)}{b} + \frac{2}{1 + u^0} \ucon{1} \ucon{a} \ucov{b};
\end{align} \end{subequations}
and
\begin{subequations} \begin{align}
  \pfrac{\source^m}{\prim^n} & =
  \begin{pmatrix}
    0 & 0 & \zerocov{b} & 0 & 0 & 0 & \zerocov{b} \\
    \frac{4}{\rho} \alphabea \Eb \ucon{0} & -\frac{4}{\pgas} \alphabea \Eb \ucon{0} & \zerocov{b} & 0 & 0 & \alphabea \ucon{0} & \alphabf \ucov{b} \\
    \frac{4}{\rho} \alphabea \Eb \ucon{a} & -\frac{4}{\pgas} \alphabea \Eb \ucon{a} & \zerosplit{a}{b} & \zerocon{a} & \zerocon{a} & \alphabea \ucon{a} & \pfrac{\source^{a+2}}{\prim^{b+8}} \\
    0 & 0 & \zerocov{b} & 0 & 0 & 0 & \zerocov{b} \\
    0 & 0 & \zerocov{b} & 0 & 0 & 0 & \zerocov{b} \\
    -\frac{4}{\rho} \alphabea \Eb \ucon{0} & \frac{4}{\pgas} \alphabea \Eb \ucon{0} & \zerocov{b} & 0 & 0 & -\alphabea \ucon{0} & -\alphabf \ucov{b} \\
    -\frac{4}{\rho} \alphabea \Eb \ucon{a} & \frac{4}{\pgas} \alphabea \Eb \ucon{a} & \zerosplit{a}{b} & \zerocon{a} & \zerocon{a} & -\alphabea \ucon{a} & \pfrac{\source^{a+8}}{\prim^{b+8}}
  \end{pmatrix}, \\
  \pfrac{\source^{a+2}}{\prim^{b+8}} & = \alphabf \kronsplit{a}{b} + \frac{\alphabf}{1 + u^0} \ucon{a} \ucov{b}, \\
  \pfrac{\source^{a+8}}{\prim^{b+8}} & = -\pfrac{\source^{a+2}}{\prim^{b+8}}.
\end{align} \end{subequations}

\section{Linear Wave Numerics}
\label{sec:linear_wave_numerics}
\setcounter{equation}{0}

In all of our linear waves, we choose $\Gammagas = 5/3$, $\alphaba = 10$ (so $\alphabea,\, \alphabpa = 10$), and $\alphabs = 10$ (so $\alphabf = 20$). With a wavelength of unity, this means we always have a total optical depth of $20$ per wavelength. All $z$-components ($u^z$, $B^z$, $F^\zb$) vanish, in both the background and the perturbation, and we take the background magnetic field to obey $B^x = B^y$. Our backgrounds always have vanishing $u^x$, $u^y$, $F^\xb$, and $F^\yb$. We normalize everything to $\rho = 1$. Table~\ref{tab:background} gives the remaining details for the background states used in Section~\ref{sec:tests:linear_waves}.

\begin{deluxetable}{ccccccc}
  \tablecaption{Linear wave background states. \label{tab:background}}
  \tablehead{\colhead{Name} & \colhead{$\prad / \pgas$} & \colhead{$\pgas$} & \colhead{$B^x$} & \colhead{$\Eb$} & \colhead{$\arad$}}
  \startdata
  H1 & 1/10 & \code{2.4976873265494906e-1} & \code{0}                     & \code{7.4930619796484720e-2} & \code{1.9253382731290976e+1} \\
  H2 & 1    & \code{2.0377358490566033e-1} & \code{0}                     & \code{6.1132075471698100e-1} & \code{3.5455008763907960e+2} \\
  H3 & 10   & \code{5.9376908979841170e-2} & \code{0}                     & \code{1.7813072693952350e+0} & \code{1.4330736087381077e+5} \\
  M1 & 1/10 & \code{1.6920473773265650e-2} & \code{1.3007872144692095e-1} & \code{5.0761421319796960e-3} & \code{6.1927521300000000e+4} \\
  M2 & 1    & \code{1.8018018018018018e-2} & \code{1.3423121104280486e-1} & \code{5.4054054054054060e-2} & \code{5.1286162500000000e+5} \\
  M3 & 10   & \code{5.1282051282051280e-2} & \code{2.2645540682891915e-1} & \code{1.5384615384615383e+0} & \code{2.2244625000000000e+5} \\
  \enddata
\end{deluxetable}

The last column is the code value of the radiation constant $\arad$. This is fixed by requiring that an intensity field in the code with $\Eb = \arad (\pgas / \rho)^4$ will be in thermal equilibrium with the gas. That is, for code unit of density $[\rho]$ (e.g., in $\g\ \cm^{-3}$), an energy density $\Eb [\rho] c^2$ will physically be in equilibrium with gas with density $\rho [\rho]$ and pressure $\pgas [\rho] c^2$. Given the physical value of $\aradzero \approx 7.6 \times 10^{-15}\ \erg\ \cm^{-3}\ \K^{-4}$, $\arad$, $[\rho]$, and the molecular weight $\mu$ are related via
\begin{equation}
  \arad = \aradzero c^6 \parenpow{\bigg}{\frac{\mu \mprot}{\kB}}{4} \frac{1}{[\rho]}.
\end{equation}

Table~\ref{tab:hydro} gives the eigenvalues and eigenvectors for the three hydrodynamical linear waves shown in Section~\ref{sec:tests:linear_waves}. For these sound waves there are no transverse perturbations:\ $u^y_0,\, F^\yb_0 = 0$. Table~\ref{tab:mhd} does the same for the five MHD linear waves studied. In all cases, the phase of the wave is shifted such that $\Im(\delta\rho_0) = 0$.

\begin{deluxetable}{cccc}
  \tablecaption{Eigenvalues and eigenvectors for hydrodynamical waves. \label{tab:hydro}}
  \tablehead{\colhead{Quantity} & \colhead{H1:\ Rightgoing Sound} & \colhead{H2:\ Rightgoing Sound} & \colhead{H3:\ Rightgoing Sound}}
  \startdata
  $\Re(\omega)$          & \code{+3.1488157526582419e+0} & \code{+3.1429599763199891e+0} & \code{+3.1422980744343483e+0} \\
  $\Im(\omega)$          & \code{-2.6190006385783764e-2} & \code{-2.2828249606164981e-2} & \code{-2.0401828576401649e-2} \\
  $\Re(\delta\rho_0)$    & \code{+8.3877889167048036e-1} & \code{+6.8142732199614686e-1} & \code{+3.7649572188184188e-1} \\
  $\Re(\delta\pgaszero)$ & \code{+3.2084488925731225e-1} & \code{+1.9049344866210005e-1} & \code{+2.9897966267824361e-2} \\
  $\Im(\delta\pgaszero)$ & \code{-9.9134535607497271e-3} & \code{-4.5406457397056052e-3} & \code{-4.3310633257571322e-4} \\
  $\Re(\delta u^x_0)$    & \code{+4.2035369927276639e-1} & \code{+3.4086195060291402e-1} & \code{+1.8829013057282032e-1} \\
  $\Im(\delta u^x_0)$    & \code{-3.4962560317947367e-3} & \code{-2.4757813488656411e-3} & \code{-1.2225011362941717e-3} \\
  $\Re(\delta\Eb_0)$     & \code{+1.2904189937790878e-1} & \code{+6.1763823482839753e-1} & \code{+9.0495607014077450e-1} \\
  $\Im(\delta\Eb_0)$     & \code{+1.5203926879090203e-3} & \code{-3.5173782805648812e-2} & \code{-4.8502044824468550e-2} \\
  $\Re(\delta F^\xb_0)$  & \code{+1.3260665610964825e-3} & \code{+2.5195284461771925e-4} & \code{-3.0510547862186877e-4} \\
  $\Im(\delta F^\xb_0)$  & \code{-6.7017329068802586e-3} & \code{-2.1004450221349644e-2} & \code{-2.4580425755539041e-2} \\
  \enddata
\end{deluxetable}

\begin{longrotatetable} \begin{deluxetable}{cccccc}
  \tablecaption{Eigenvalues and eigenvectors for MHD waves. \label{tab:mhd}}
  \tablehead{\colhead{Quantity} & \colhead{M1:\ Rightgoing Fast} & \colhead{M2:\ Entropy} & \colhead{M2:\ Rightgoing Slow} & \colhead{M2:\ Rightgoing Fast} & \colhead{M3:\ Rightgoing Fast}}
  \startdata
  $\Re(\omega)$          & \code{+1.4030907769530285e+0} & \code{0}                      & \code{+6.3852389015453093e-1} & \code{+1.5949141818449315e+0} & \code{+3.1544115249869811e+0} \\
  $\Im(\omega)$          & \code{-2.3261803772811441e-2} & \code{-2.2389397045278464e-1} & \code{-6.4155513213079954e-2} & \code{-8.0455282835428230e-2} & \code{-1.8555581698888605e-2} \\
  $\delta\rho_0$         & \code{+9.5533821762932847e-1} & \code{+9.9745867367582863e-1} & \code{+9.3698871700076369e-1} & \code{+9.5003889374992712e-1} & \code{+4.2116116065164089e-1} \\
  $\Re(\delta\pgaszero)$ & \code{+2.4545590135418679e-2} & \code{+1.3833062172065714e-2} & \code{+2.0454458269493797e-2} & \code{+2.3093435201993501e-2} & \code{+2.8884767730924427e-2} \\
  $\Im(\delta\pgaszero)$ & \code{-1.7212846538958446e-3} & \code{0}                      & \code{-3.5293728455285493e-3} & \code{-1.9397668089885904e-3} & \code{-4.0832850598264241e-4} \\
  $\Re(\delta u^x_0)$    & \code{+2.1333546226860403e-1} & \code{0}                      & \code{+9.5220760069989752e-2} & \code{+2.4115642478578933e-1} & \code{+2.1143982774443940e-1} \\
  $\Im(\delta u^x_0)$    & \code{-3.5368828179817360e-3} & \code{-3.5543274930418547e-2} & \code{-9.5672798230796927e-3} & \code{-1.2165111191924344e-2} & \code{-1.2437784249231532e-3} \\
  $\Re(\delta u^y_0)$    & \code{-9.4540134499253400e-2} & \code{0}                      & \code{+2.2570187473805797e-1} & \code{-7.3220148652734046e-2} & \code{-1.0692349015159086e-2} \\
  $\Im(\delta u^y_0)$    & \code{-3.1336310080776267e-3} & \code{-3.2950611052758337e-2} & \code{-1.0824232344232039e-1} & \code{-6.1181702675864730e-3} & \code{+1.0200405321943583e-3} \\
  $\Re(\delta B^y_0)$    & \code{+1.7929397695717086e-1} & \code{+9.7664604496020095e-3} & \code{-1.8358914684765970e-1} & \code{+1.6598302737695866e-1} & \code{+1.0019976240705387e-1} \\
  $\Im(\delta B^y_0)$    & \code{+2.7376133858246950e-3} & \code{0}                      & \code{+1.1188967996089302e-1} & \code{+5.1753405380474959e-3} & \code{-4.3172409871501671e-4} \\
  $\Re(\delta\Eb_0)$     & \code{+9.7030693142753782e-3} & \code{-4.9128596680383302e-2} & \code{+4.2597245178004423e-2} & \code{+7.1308949711257980e-2} & \code{+8.7423772851515402e-1} \\
  $\Im(\delta\Eb_0)$     & \code{-1.5553274585766021e-3} & \code{0}                      & \code{-4.1582610315888224e-2} & \code{-2.1953236496559857e-2} & \code{-4.5633199309910999e-2} \\
  $\Re(\delta F^\xb_0)$  & \code{-9.4902938599335495e-5} & \code{0}                      & \code{-4.1837076267764771e-3} & \code{-1.6679522286032048e-3} & \code{-3.1252339580678350e-4} \\
  $\Im(\delta F^\xb_0)$  & \code{-9.2256600777633675e-4} & \code{+5.1739785366057365e-3} & \code{-4.3915401620257541e-3} & \code{-6.2430775123552551e-3} & \code{-2.3216160870051111e-2} \\
  $\Re(\delta F^\yb_0)$  & \code{+3.8836674694259790e-6} & \code{0}                      & \code{+2.8603718033243122e-4} & \code{+4.7659405424480282e-5} & \code{+1.9086396938480885e-4} \\
  $\Im(\delta F^\yb_0)$  & \code{-4.4693666195008915e-5} & \code{-2.6886365431057796e-5} & \code{+5.0506463526206967e-4} & \code{-4.2049304677408794e-4} & \code{-3.4304273514267473e-3} \\
  \enddata
\end{deluxetable} \end{longrotatetable}

\end{CJK*}
\end{document}